# Unification and Matching on Compressed Terms

ADRIÀ GASCÓN and GUILLEM GODOY
Universitat Politècnica de Catalunya
and
MANFRED SCHMIDT-SCHAUSS
Inst. f. Informatik, Goethe-Universität

Term unification plays an important role in many areas of computer science, especially in those related to logic. The universal mechanism of grammar-based compression for terms, in particular the so-called *Singleton Tree Grammars (STG)*, have recently drawn considerable attention. Using STGs, terms of exponential size and height can be represented in linear space. Furthermore, the term representation by directed acyclic graphs (dags) can be efficiently simulated. The present paper is the result of an investigation on term unification and matching when the terms given as input are represented using different compression mechanisms for terms such as dags and Singleton Tree Grammars. We describe a polynomial time algorithm for context matching with dags, when the number of different context variables is fixed for the problem. For the same problem, NP-completeness is obtained when the terms are represented using the more general formalism of Singleton Tree Grammars. For first-order unification and matching polynomial time algorithms are presented, each of them improving previous results for those problems.

Categories and Subject Descriptors: F.4.1 [**Theory of computation**]: Mathematical Logic and formal languages—*lambda calculus and related systems*; F.4.2 [**Theory of computation**]: Mathematical Logic and formal languages—*Grammars and Other Rewriting Systems*

## 1. INTRODUCTION

The task of solving equations is an important component of any mathematically founded science. In general, solving an equation $s \doteq t$ consists of finding a substitution $\sigma$ for variables occurring in both expressions $s$ and $t$ such that $\sigma(s) = \sigma(t)$ holds. The range for the variables, the kind of expressions $s$ and $t$, and their semantics, as well as the semantics of $=$ depend on the context. By specifying some of these parameters we can define the well-known first-order term unification problem. In the context of this problem the expressions $s$ and $t$ are terms with leaf variables standing for terms, all function symbols are non-interpreted, and $=$ is interpreted as syntactic equality.

The term matching problem is a particular case of term unification. It is characterized by the condition that one of the sides of the equation $s \doteq t$, say $t$, contains no variables. Like term unification, this is a common problem in areas like functional and logic programming, automated deduction, deductive databases, artificial intelligence, information retrieval, compiler design, type checking in programming languages, etc.

The first-order term unification and matching problems are efficiently solvable and there is a history of different algorithms: decidable, but exponential [Robinson 1965]; linear time, but using a very special term representation [Paterson and Wegman 1978]; and an almost linear one, using a variant of term compression: [Martelli and Montanari 1982]. The expressivity of first-order terms is often insufficient to deal with some of the current challenges in the mentioned areas. This motivates the study of some variants and generalizations of the first-order term matching and unification problems. In this sense, incorporating more complex interpretations of the function symbols and equality predicates under equational theories has been widely considered (see [Baader and Siekmann 1994; Baader and Snyder 2001a]). Further extensions like allowing other kinds of variables related to terms have also been explored. This is the case of context variables, i.e. variables that can be substituted by contexts, which are trees with a single hole (syntactically, the hole is a special constant). For example, consider the term $t = f(g(a,b), g(a,h(b)))$. Then the match-equation $F(a) \doteq t$, where $F$ is a context variable, has the solutions $F \mapsto f(g([\cdot],b), g(a,h(b)))$ (where $[\cdot]$ means the hole) and $F \mapsto f(g(a,b), g([\cdot], h(b)))$; the equation $f(F(b), F(h(b))) \doteq t$ has the solution $F \mapsto g(a, [\cdot])$, whereas $f(F(b), F(b)) \doteq t$ has no solution. Context matching is known to be NP-complete, but there are several subcases that can be solved efficiently [Schmidt-Schauß and Stuber 2004].

As illustrated by the example above, the instantiation of a context variable by a match is a context, i.e. a tree with a hole. Thus multiple occurrences of the same context variable correspond to the question whether there are occurrences of the same subtree, but up to one position in the subtree. This has applications in computational linguistics [Niehren et al. 1997]. It is also easy to encode questions that ask for subtrees that are equal up to several positions. Some applications of context matching are in querying XML-data bases: see [Berglund et al. 2007] for the XPATH-standard, [Schmidt-Schauß and Stuber 2004] for investigating context matching, and [Gottlob et al. 2006] for analyzing conjunctive query mechanisms over trees. Another interesting application of context matching is the search within tree structures and the corresponding extraction of information. For example, the match equation $F(s) = t$ where $t$ is ground and $s$ has no occurrences of $F$ corresponds to the question whether there is a subtree of $t$ that is matched by $s$. This can easily be combined as conjunctive search $F_1(s_1) = t; \ldots; F_n(s_n) = t$, where the $F_i$ are pairwise different and do not occur elsewhere. These match equations correspond to the search question whether there are subterms $r_i$ of $t$ that can be matched by $s_i$ for $i = 1, \ldots, n$, where variables within $s_i$ must have a common instance in $t$.

Besides adding expressiveness to these problems it is also necessary to take account of the feasibility of implementing the algorithms to which we refer. In that sense, an option is to reconsider complexity issues for the original problems or its variants by assuming that the input terms are given in some compressed representation. Many of the applications dealing with the problems we have introduced and their variants require some kind of internal succinct representation for terms, in order to guarantee computability in an environment with a limited amount of resources. It is well-known that first-order unification may require exponential space with a plain term representation whereas only polynomial space is required when dags are used for representing terms. Similarly, if terms are large but have lots of common subterms, like $t_1 = f(a,b), t_2 = f(t_1,t_1), \ldots, t_n = f(t_{n-1},t_{n-1})$, then the context matching equation $F(a) = t_n$ requires exponential space using the plain term representation to represent $t_n$, whereas a dag representation requires linear space. This motivates to investigate context matching with compression techniques like dags. Although the context matching problem is NP-complete, sometimes it suffices to consider a small number of context variables, which can be thought of as fixed for the problem. This kind of restriction has already been considered for context unification restricted to two context variables [Schmidt-Schauß and Schulz 2002], and also proved useful in the context of program verification with procedure calls [Gulwani and Tiwari 2007; Gascón et al. 2009], where context unification for (monadic and multi-ary, respectively) signatures and a single context variable allows the automatic generation of invariants.

Besides the dag representation, more general grammar-based compression mechanisms for terms have recently drawn considerable attention in research. In particular, a *Singleton Tree Grammars (STG)* can succinctly represent terms/trees which are exponentially big in size and height. Grammar-based compression techniques were initially applied to words [Plandowski 1994; Plandowski and Rytter 1999] and led to important results in string processing, with applications [Hirao et al. 2000; Genest and Muscholl 2002; Lasota and Rytter 2006] in software/hardware verification, information retrieval, and bioinformatics. Efficient algorithms have been developed for checking whether two compressed inputs represent the same word/term [Plandowski 1995; Lohrey 2006; Lifshits 2007], also randomized algorithms for the equality test [Gasieniec et al. 1996; Berman et al. 2002; Schmidt-Schauß and Schnitger 2009], and for finding occurrences of one of them within the other (fully compressed pattern matching) [Karpinski et al. 1995; Karpinski et al. 1996; Miyazaki et al. 1997; Lifshits 2007]. In that sense, *Straight-Line Programs (SLP)*, or the equivalent formalism of *Singleton Context Free Grammars (SCFG)* are now a widely accepted formalism for text compression. Essentially, an SCFG, i.e. a context free grammar where all non-terminals generate a singleton language, is used for representing single words. This notion was extended from words to terms [Busatto et al. 2005; Schmidt-Schauß 2005; Comon et al. 1997] such that every non-terminal in a *Singleton Tree Grammar (STG)* represents exactly one tree. It led to applications in XML tree structure compression [Busatto et al. 2005] and XPATH [Lohrey and Maneth 2005]. STGs have also been proved useful for complexity analysis of unification algorithms in [Levy et al. 2006b; 2006a]. Recently, it was shown that tree grammars using multi-hole-contexts are polynomially equivalent to STGs [Lohrey et al. 2009]. Moreover, STG-based compressors have already been developed [Maneth et al. 2008].

Our focus is not on how to compress terms (which can be found e.g. in [Schmidt-Schauß 2005; Busatto et al. 2008] and for deduction system in [Cheney 1998; Graf 1995; 1996]), but on efficient algorithms on compressed terms. This paper is an extended and improved version of two conference papers [Gascón et al. 2009; 2008] by the same authors. In [Gascón et al. 2008], the context matching problem for a fixed number of context variables where the input terms are represented with dags was proven decidable in polynomial time. In our view this result shows that compression is well-behaved for context matching and should be considered. Furthermore, NP-completeness was shown for the context matching problem with terms compressed using STGs. In the present paper we improve the description of both algorithms with additional remarks and a more rigorous notation (by representing dags as a particular case of STGs). This change in notation allows to be more precise in explanations, proofs, and even in the complexity analysis. These two results are presented in Section 3, and Section 5, respectively, where it is shown that context matching with dags with $k$ context variables can be solved in time $\mathcal{O}((\texttt{depth}(G))^{k+1}|G|^2\texttt{log}(|G|))$, where $|G|$ is the size of the initial dag $G$ (see Theorem 3.19). As a complement we prove that context matching with STG-compressed terms is NP-complete (see Theorem 5.10). Section 4 contains several technical algorithms and constructions on SCFGs and STGs, which are indispensable for showing polynomial space and/or time behavior of the matching and unification algorithms. Also in [Gascón et al. 2008], and in [Gascón et al. 2009], there were polynomial time algorithms presented for the first-order matching and first-order unification problems, in both cases with terms represented with STGs. As a novel contribution we describe, in Section 6 and Section 7, faster algorithms for these two problems: The first-order unification algorithm on STG-compressed terms runs in time $\mathcal{O}(|V|(|G|+|V|\texttt{depth}(G))^3)$, where $V$ is the set of variables, and $G$ is the input STG (see Theorem 6.2) and the matching algorithm in time $\mathcal{O}((|G|+|V|\texttt{depth}(G))^3$, (see Theorem 7.3). Moreover, we believe that the presented solutions are also a gain in simplicity which makes them easily implementable.

## 2. PRELIMINARIES

A *signature* is a set $\mathcal{F}$ together with a function $\texttt{ar}: \mathcal{F} \to \mathbb{N}$. Members of $\mathcal{F}$ are called function symbols, and $\texttt{ar}(f)$ is called the *arity* of the function symbol $f$. Function symbols of arity 0 are called constants. Let $\mathcal{X}$ be a set disjoint from $\mathcal{F}$ whose elements are called variables. We assume the function $\texttt{ar}$ to be also defined for variables, i.e. $\texttt{ar}: (\mathcal{F} \cup \mathcal{X}) \to \mathbb{N}$, but with $\texttt{ar}(V) \in \{0,1\}$ for variables $V \in \mathcal{X}$. Variables with arity 0, denoted $x, y, z$ with possible indexes, are called first-order variables, and variables with arity 1, denoted $F$ with possible subscripts, are called context variables. We use $f, g$, with possible indexes, for denoting an element of $\mathcal{F}$, and $\alpha$ for denoting an element in $\mathcal{F} \cup \mathcal{X}$.

The set $\mathcal{T}(\mathcal{F} \cup \mathcal{X})$ of terms over $\mathcal{F}$ and $\mathcal{X}$, also denoted $\mathcal{T}(\mathcal{F}, \mathcal{X})$, is defined to be the smallest set having the property that $\alpha(t_1, \ldots, t_m) \in \mathcal{T}(\mathcal{F} \cup \mathcal{X})$ whenever $\alpha \in (\mathcal{F} \cup \mathcal{X})$, $m = \texttt{ar}(\alpha)$ and $t_1, \ldots, t_m \in \mathcal{T}(\mathcal{F} \cup \mathcal{X})$. The set $\mathcal{T}(\mathcal{F})$ is called the set of ground terms over $\mathcal{F}$, that is, the subset of terms of $\mathcal{T}(\mathcal{F} \cup \mathcal{X})$ with no occurrences of variables. We denote by $s, t$, with possible indexes, terms in $\mathcal{T}(\mathcal{F} \cup \mathcal{X})$.

The size $|t|$ of a term $t$ is the number of occurrences of variables and function symbols in $t$. The *height* of a term $t$, denoted $\texttt{height}(t)$, is 0 if $t$ is a constant or a first-order variable, and $1 + \max\{\texttt{height}(t_1), \ldots, \texttt{height}(t_m)\}$ if $t = \alpha(t_1, \ldots, t_m)$. *Positions* of a term $t$, denoted $p, q$ with possible subindexes, are sequences of natural numbers that are used to identify the location of subterms of $t$. The set $Pos(t)$ of *positions* of $t$ is defined by $Pos(t) = \{\lambda\}$ if $t$ is a constant or a variable, and $Pos(t) = \{\lambda\} \cup \{1 \cdot p \mid p \in Pos(t_1)\} \cup \ldots \cup \{n \cdot p \mid p \in Pos(t_m)\}$ if $t = \alpha(t_1, \ldots, t_m)$, where $\lambda$ denotes the empty sequence and $p \cdot q$, or simply $pq$, denotes the concatenation of $p$ and $q$. If $t$ is a term and $p$ a position, then $t|_p$ is the subterm of $t$ at position $p$. More formally defined, $t|_\lambda = t$ and $\alpha(t_1, \ldots, t_m)|_{i \cdot p} = t_i|_p$. We can define a partial order $\preceq$ on $Pos(t)$ by $p \preceq q$ if and only if $p$ is a prefix of $q$, i.e. there is a sequence $p'$ such that $q = p \cdot p'$. We say that positions $p$ and $q$ are *disjoint* if they are incomparable with respect to $\preceq$. We denote by $\texttt{pre}(t)$ the preorder traversal (as a word) of a term $t$. It is recursively defined as $\texttt{pre}(t) = t$, if $t$ has arity 0, and $\texttt{pre}(t) = \alpha \cdot \texttt{pre}(t_1) \cdot \ldots \cdot \texttt{pre}(t_m)$, if $t = \alpha(t_1, \ldots, t_m)$. Two arbitrary different trees may have the same preorder traversal, but when they represent terms over a fixed signature where the arity of every function symbol is fixed, the preorder traversal is unique for every term. Given a term $t$, there is a natural bijective mapping between the indexes $\{1, \ldots, |\texttt{pre}(t)|\}$ of $\texttt{pre}(t)$ and the positions $Pos(t)$ of $t$, which associates every position $p \in Pos(t)$ to the index $i \in \{1, \ldots, |\texttt{pre}(t)|\}$ you find at $\texttt{root}(t|p)$ while traversing the tree in preorder. We can recursively define the two mappings $\texttt{pIndex}(t, p) \to \{1, \ldots, |\texttt{pre}(t)|\}$ and $\texttt{iPos}(t, i) \to Pos(t)$ as follows. $\texttt{pIndex}(t, \lambda) = 1$, $\texttt{pIndex}(\alpha(t_1, \ldots, t_m), i.p) = (1 + |t_1| + \ldots + |t_{i-1}|) + \texttt{pIndex}(t_i, p)$, $\texttt{iPos}(t, 1) = \lambda$, and $\texttt{iPos}(\alpha(t_1, \ldots, t_m), 1 + |t_1| + \ldots + |t_{i-1}| + k) = i.\texttt{iPos}(t_i, k)$ for $1 \leq k \leq |t_i|$.

Intuitively, contexts are terms with a single occurrence of a hole $[\cdot]$ into which terms (or other contexts) may be inserted. We denote contexts by upper case letters $C, D$. The set of contexts over $\mathcal{F}$ and $\mathcal{X}$ is denoted by $\mathcal{C}(\mathcal{F} \cup \mathcal{X})$ whereas the set of ground contexts over $\mathcal{F}$ is denoted $\mathcal{C}(\mathcal{F})$. We can provide a formal definition by considering a context to be a term in an extended signature that includes an extra constant symbol $[\cdot]$, and where this symbol occurs exactly once in the term. Hence, the smallest context contains just the hole and has size 1. If $C$ and $D$ are contexts and $s$ is a term, $CD$ and $Cs$ represent the context and the term that are like $C$ except that the occurrence of $[\cdot]$ is replaced by $D$ and $s$, respectively. If $D_1 = D_2 D_3$ for contexts $D_1, D_2, D_3$, then $D_2$ is called a *prefix* of $D_1$, and $D_3$ is called a *suffix* of $D_1$. The position of the hole in a context $C$ is called *hole path*, and denoted $\texttt{hp}(C)$, and its length is denoted as $|\texttt{hp}(C)|$.

A *substitution* is a mapping $\mathcal{X} \to \mathcal{T}(\mathcal{F}, \mathcal{X}) \cup \mathcal{C}(\mathcal{F}, \mathcal{X})$ relating first-order variables to terms, and context variables to contexts. Substitutions can also be applied to arbitrary terms by homomorphically extending them by $\sigma(f(t_1, \ldots, t_m)) = f(\sigma(t_1), \ldots, \sigma(t_m))$ and $\sigma(F(t)) = \sigma(F)\sigma(t)$.

An instance of the *context unification problem* is a set of equations $\Delta = \{s_1 \doteq t_1, \ldots, s_n \doteq t_n\}$, where the $s_i, t_i$ are terms in $\mathcal{T}(\mathcal{F}, \mathcal{X})$. The question is to compute a substitution $\sigma$ (the solution), such that $\sigma(s_i) = \sigma(t_i)$ for all $i$. The *context matching problem* is a particular case of context unification where one of the sides of each equation in $\Delta$ is ground.

With $[i, n]$ we denote the set $\{i, i+1, \ldots, n\} \subseteq \mathbb{N}$.

Unification and Matching on Compressed Terms · 5

2.1 Compressed term representation

*Definition* 2.1. A *singleton context-free grammar (SCFG)* $G$ is a 3-tuple $\langle \mathcal{N}, \Sigma, R \rangle$, where $\mathcal{N}$ is a set of non-terminals, $\Sigma$ is a set of symbols (a signature), and $R$ is a set of rules of the form $N \to \alpha$ where $N \in \mathcal{N}$ and $\alpha \in (\mathcal{N} \cup \Sigma)^*$. The sets $\mathcal{N}$ and $\Sigma$ must be disjoint, each non-terminal $X$ appears as a left-hand side of just one rule of $R$, and there exists a well founded ordering $>_G$ such that $X \to r_1 Y r_2 \in R$ implies $X >_G Y$ for any $X, Y \in \mathcal{N}$. The word generated by a non-terminal $N$ of $G$, denoted by $w_{G,N}$ or $w_N$ when $G$ is clear from the context, is the word in $\Sigma^*$ reached from $N$ by successive applications of the rules of $G$.

*Definition* 2.2. A *singleton tree grammar (STG)* is a 4-tuple $G = (\mathcal{TN}, \mathcal{CN}, \Sigma, R)$, where $\mathcal{TN}$ are tree/term non-terminals, or non-terminals of arity 0, $\mathcal{CN}$ are context non-terminals, or non-terminals of arity 1, and $\Sigma$ is a signature of function symbols (the terminals), such that the sets $\mathcal{TN}$, $\mathcal{CN}$, and $\Sigma$ are pairwise disjoint. The set of non-terminals $\mathcal{N}$ is defined as $\mathcal{N} = \mathcal{TN} \cup \mathcal{CN}$. The rules in $R$ may be of the form:

—$A \to \alpha(A_1, \ldots, A_m)$, where $A, A_i \in \mathcal{TN}$, and $\alpha \in \Sigma$ is an $m$-ary terminal symbol.
—$A \to C_1 A_2$ where $A, A_2 \in \mathcal{TN}$, and $C_1 \in \mathcal{CN}$.
—$C \to [\cdot]$ where $C \in \mathcal{CN}$.
—$C \to C_1 C_2$, where $C, C_1, C_2 \in \mathcal{CN}$.
—$C \to \alpha(A_1, \ldots, A_{i-1}, C_i, A_{i+1}, \ldots, A_m)$, where $A_1, \ldots, A_{i-1}, A_{i+1}, \ldots, A_m \in \mathcal{TN}$, $C, C_i \in \mathcal{CN}$, and $\alpha \in \Sigma$ is an $m$-ary terminal symbol.
—$A \to A_1$, ($\lambda$-rule) where $A$ and $A_1$ are term non-terminals.

Let $N_1 >_G N_2$ for two non-terminals $N_1, N_2$, iff $(N_1 \to t)$, and $N_2$ occurs in $t$. The STG must be non-recursive, i.e. the transitive closure $>_G^+$ must be terminating. Furthermore, for every non-terminal $N$ of $G$ there is exactly one rule having $N$ as left-hand side. Sometimes we refer to the right-hand side of this rule as the *definition* of $N$ in $G$. Given a term $t$ with occurrences of non-terminals, the derivation of $t$ by $G$ is an exhaustive iterated replacement of the non-terminals by the corresponding right-hand sides. The result is denoted as $w_{G,t}$. In the case of a non-terminal $N$ we also say that $N$ *generates* $w_{G,N}$. We will write $w_N$ when $G$ is clear from the context.

Note that we have used $\Sigma$ instead of $\mathcal{F}$ for denoting the set of terminals of the grammar, although it is also a signature. We explain the reasons as follows. In this paper, STGs are used for representing terms and contexts. In particular, a terminal $A$ of an STG $G$ generates a term. If $\Sigma$ was $\mathcal{F}$ we would be able to represent just ground terms. But we want to represent non-ground terms, i.e. terms with occurrences of first-order and context variables. Thus, $\Sigma$ must also contain variables, of arity 0 if they are first-order variables, and of arity 1 if they are context variables. We will represent a substitution application $\{V \to t\}$ by converting the variable $V$ from a terminal into a non-terminal of the grammar and adding the necessary rules such that it generates $t$. Thus, in this setting, variables can be represented both by terminals and non-terminals of the grammar.

Given an STG $G = (\mathcal{TN}, \mathcal{CN}, \Sigma, R)$ we can refer to the set $\mathcal{T}(\mathcal{TN} \cup \mathcal{CN} \cup \Sigma)$ of terms over the terminals and non-terminals of $G$ where symbols in $\mathcal{TN}$ have arity 0

ACM Transactions on Computational Logic, Vol. 0, No. 0, March 2010.

and symbols in $\mathcal{CN}$ have arity 1. Similarly, we can refer to the set $\mathcal{C}(\mathcal{TN} \cup \mathcal{CN} \cup \Sigma)$ of contexts over the terminals and non-terminals of $G$.

With respect to the notation used in this paper, we denote indiscriminately terms in $\mathcal{T}(\mathcal{TN} \cup \mathcal{CN} \cup \Sigma)$, and $\mathcal{T}(\mathcal{F}, \mathcal{X})$ by $s, t, u, v$, with possible indexes, since at each point of this paper it is clear from the context to which set we refer. By capital letters $A, B$ we refer to term non-terminals and by $C, D$ we refer to context non-terminals of a given STG. By $N$ we denote a non-terminal of the grammar in general. We denote by $\alpha$ the terminals of the grammar in general, by $f, g$ the terminals of the grammar which represent a function symbol, by $F$, with possible indexes, both the terminals and non-terminals of the grammar representing context variables, and finally, we denote by $x, y, z$ both the terminals and non-terminals of the grammar representing first-order variables.

Now that the set $\mathcal{T}(\mathcal{TN} \cup \mathcal{CN} \cup \Sigma)$ has been introduced, given a term $t \in \mathcal{T}(\mathcal{TN} \cup \mathcal{CN} \cup \Sigma)$, we can define $w_{G,t}$ more formally.

*Definition* 2.3. Let $G = (\mathcal{TN}, \mathcal{CN}, \Sigma, R)$ be an STG. Let $t$ be a term in $\mathcal{T}(\mathcal{TN} \cup \mathcal{CN} \cup \Sigma)$ or a context in $\mathcal{C}(\mathcal{TN} \cup \mathcal{CN} \cup \Sigma)$. Then, we define $w_t$ recursively as follows.

—If $t = \alpha(t_1, \ldots, t_n)$ for some terminal $\alpha \in \Sigma$ of arity $m$ then $w_t = \alpha(w_{t_1}, \ldots, w_{t_n})$.
—If $t = N$ for some non-terminal $N$ of $G$ with a rule $N \to u \in R$ then $w_t = w_u$.
—If $t = C(t_1)$ for some context non-terminal $C$ of $G$ then $w_t = w_C w_{t_1}$.
—If $t = [\cdot]$ then $w_t = [\cdot]$.

*Definition* 2.4. Let $G = (\mathcal{TN}, \mathcal{CN}, \Sigma, R)$ be an STG. Let $S$ be a set of non-terminals of $G$. We define $\texttt{restriction}(G, S) = (\mathcal{TN}', \mathcal{CN}', \Sigma, R')$ as the STG where $\mathcal{TN}' \subseteq \mathcal{TN}$, $\mathcal{CN}' \subseteq \mathcal{CN}$, and $R' \subseteq R$ are the smallest sets such that $G' = \texttt{restriction}(G, S)$ satisfies $w_{G',N} = w_{G,N}$ for each non-terminal $N$ in $S$.

A directed acyclic graph (dag) can be defined as a particular case of an STG (in fact, this representation is in direct correspondence with the classic implementation of dags using adjacency lists).

*Definition* 2.5. A *DAG* is an STG where the set of context non-terminals $\mathcal{CN}$ is empty, and moreover, there are only rules of the form $A \to f(A_1, \ldots, A_m)$.

*Example* 2.6. The STG $\{A_0 \to f(A_1, A_1),\ A_1 \to f(A_2, A_2),\ \ldots,\ A_{n-1} \to f(A_n, A_n)\ A_n \to a\}$ is a DAG that represents the complete binary tree of height $n$ over a function symbol $f$ and a constant $a$. The size of this term is exponential, whereas its height is linear.

Nevertheless, STG-represented terms may have exponential height in the size of the grammar in contrast to dags, which only allow for a linear height in the (notational) size of the dags.

*Example* 2.7. The STG $\{C_0 \to C_1 C_1,\ C_1 \to C_2 C_2,\ C_2 \to C_3 C_3,\ \ldots,\ C_{n-1} \to C_n C_n,\ C_n \to g(C),\ C \to [\cdot]\}$ represents the context $w_{C_0} = g^{2^n}[\cdot]$, whose height is exponential. This is not a DAG.

A DAG $G$ is called *optimally compressed* if equal terms are represented by the same term non-terminal. The test whether a DAG is optimally compressed can be performed in time $\mathcal{O}(n \cdot \log n)$, and a transformation into optimally compressed form in time $\mathcal{O}(n \cdot \log n)$.

*Definition* 2.8. The *size* $|G|$ of an STG $G$ is the sum of the sizes of its rules, where the size of a rule $N \to u$ is $1 + |u|$. The *depth* within $G$ of a non-terminal $N$ is defined recursively as $\texttt{depth}(N) := 1 + \max\{\texttt{depth}(N') \mid N'$ is a non-terminal in $u$ where $N \to u \in G\}$ and the maximum of an empty set is assumed to be 0. The *depth of a grammar* $G$ is the maximum of the depths of all non-terminals of $G$, and it is denoted as $\texttt{depth}(G)$.

If the signature is fixed, then we could also use the number of rules as a complexity measure of STGs.

Plandowski [Plandowski 1994; 1995] proved decidability in polynomial time for the word problem for SCFGs, i.e., given an SCFG $P$ and two non-terminals $A$ and $B$, to decide whether $w_A = w_B$. The best complexity for this problem has been obtained recently by Lifshits [Lifshits 2007] with time $\mathcal{O}(|P|^3)$. In [Busatto et al. 2005; Schmidt-Schauß 2005; Busatto et al. 2008] Plandowski's result is generalized to STGs. Since the result in [Busatto et al. 2005] is based on a linear reduction from terms to words and a direct application of Plandowski's result, it also holds for the Lifshits result. Hence, we have the following.

THEOREM 2.9. *([Lifshits 2007; Busatto et al. 2005; 2008]) Given an STG $G$, and two tree non-terminals $A, B$ from $G$, it is decidable in time $\mathcal{O}(|G|^3)$ whether $w_A = w_B$.*

Several properties of STGs are efficiently decidable. The following lemmas will be used all along the paper.

LEMMA 2.10. *Let $G$ be an STG. The number $|w_N|$, for every non-terminal $N$ of $G$, is computable in time $\mathcal{O}(|G|)$.*

PROOF. We give an alternative definition of $|w_N|$ recursively as follows.

—if $(N \to f(N_1, \ldots, N_m)) \in G$ then $|w_N| = 1 + |w_{N_1}| + \ldots + |w_{N_m}|$, where $N_1, \ldots, N_m$ are non-terminals of $G$ and $f$ is a function symbol with $\texttt{ar}(f) = m$.
—if $N \to C_1 N_2$ then $|w_N| = |w_{C_1}| + |w_{N_2}| - 1$, where $C_1$ is a context non-terminal and $N_2$ is a non-terminal of $G$.

The correctness of the above definition can be shown by induction on the size of $w_N$. Moreover, since the recursive calls in the definition of $|w_N|$ will be done, at most, over all the non-terminals of $G$, $|w_N|$ is computable in linear time over $|G|$ using a dynamic programming scheme. □

LEMMA 2.11. *Given an STG $G$, a terminal $\alpha$, and a non-terminal $N$ of $G$, it is decidable in time $\mathcal{O}(|G|)$ whether $\alpha$ occurs in $w_N$.*

PROOF. Whether $\alpha$ occurs in $w_N$ can be computed efficiently again using a dynamic programming scheme: note that $\alpha$ occurs in $w_N$ iff either $w_N \to \alpha \in G$, or $\alpha$ occurs in $w_{N'}$ for some non-terminal $N'$ occurring in the right-hand side of the rule for $N$. □

## 3. A PTIME ALGORITHM FOR $K$-CONTEXT MATCHING WITH DAGS

The context matching problem is NP-complete [Schmidt-Schauß and Schulz 1998]. In this section we reconsider this problem by introducing the additional restriction stating that the maximum number $k$ of different context variables of a given instance is fixed for the problem. We refer to this problem as *k-context matching*, which is in fact a family of problems indexed by $k$. Our goal is to prove that a complete representation of all solutions is computable in polynomial time when the input terms are represented with dags. This variant is called *k-context matching with dags* (k-CMD problem).

Our algorithm is presented as non-deterministic, but where the guessing can choose only from a polynomial number of possibilities. In Subsection 3.1, we solve the problem for the simpler case of uncompressed terms. This case is easy, but serves for a better understanding of some ideas appearing later, and shows the use of the non-determinism for simplifying explanations. In Subsection 3.2, we explain a situation where the context solution for a context variable can be inferred. It is used several times in the algorithm. In Subsection 3.3, we give the intuition behind the algorithm in order to help understanding the technical difficulties. In Subsection 3.4 we specify the data representation used in the algorithm, based on STGs. We explain the advantages of using STGs for representing dags, such as clarity, but also simplicity when analyzing complexity of the required operations for this problem. In Subsection 3.5 we present the set of rules of the algorithm, prove that they are sound and complete, and that they give in fact a complete representation of all the solutions for the initial set of equations. In Subsection 3.6 we analyze complexity issues.

### 3.1 k-Context Matching for Uncompressed Terms

A non-deterministic polynomial time algorithm with few guessings can be easily obtained for the $k$-context matching problem. Suppose we are given an instance $\{s \doteq t\}$ of the problem, where $t$ is a ground term and $s$ contains at most $k$ different context variables. Any solution of $\{s \doteq t\}$ instantiates every context variable by a context occurring in $t$. The number of different contexts in $t$ is bounded by $|t|^2$. This is because any context occurring in $t$ can be defined by two positions of $t$: the root position and the hole position of the context. Hence, it suffices to do at most $k$ guessings of contexts for the context variables among $|t|^2$ possibilities. After applying this partial substitution, we have to check if the resulting first-order matching problem has a solution. Since $k$ is assumed to be fixed, the overall execution time is polynomial.

When the input is compressed with dags, the problem becomes more difficult. In particular, the number of different contexts of the right-hand side can be exponential in the size of the input. For example, $t_1$ defined by $t_1 = f(t_2, t'_2), t_2 = f(t_3, t'_3), t'_2 = f(t'_3, t_3), \ldots, t_n = f(a, b), t'_n = f(b, a)$ has $2^{n-1}$ different contexts with the argument $a$, which precludes an efficient test for all contexts, e.g. in the matching problem $F(f(f(b, a), f(a, b))) \doteq t_1$.

### 3.2 Inferring the Joint Context

One of the key points for obtaining a polynomial time algorithm is the fact that in some cases, the (joint) context solution for a context variable can be inferred. Consider the simple case where we have two matching equations of the form $F(s) \doteq u$ and $F(t) \doteq v$, and suppose that $u$ and $v$ are different. Suppose also that we know the existence of a solution $\sigma$ for these equations, but the only known information for $\sigma$ is $|\mathtt{hp}(\sigma(F))|$, i.e. just the length of the hole position of $\sigma(F)$ and nothing else. It can be proved that this information suffices to obtain $\sigma(F)$. With this aim we define below $\mathtt{JointCon}(u,v,l)$ for any terms $u$ and $v$, and natural number $l$, which intuitively corresponds to the supposed $|\mathtt{hp}(\sigma(F))|$.

*Definition* 3.1. Let $u \neq v$ be terms, let $l \in \mathbb{N}$. We define $\mathtt{JointCon}(u,v,0)$ to be the empty context $[\cdot]$. We also define $\mathtt{JointCon}(f(u_1,\ldots,u_m),g(v_1,\ldots,v_m),l+1) = f(u_1,\ldots,u_{i-1},\mathtt{JointCon}(u_i,v_i,l),u_{i+1},\ldots,u_m)$ in the case where $f = g$ and there exists $i \in [1,m]$ such that $u_j = v_j$ for all $j \neq i$. Otherwise, $\mathtt{JointCon}(f(u_1,\ldots,u_m),f(v_1,\ldots,v_m),l+1)$ is undefined.

Note that in the second case of the previous definition, if $f = g$ and such an $i$ exists, then it is unique. This is because $f(u_1,\ldots,u_m)$ and $g(v_1,\ldots,v_m)$ are different, and hence, $u_j = v_j$ for all $j \neq i$ implies that $u_i \neq v_i$.

*Example* 3.2. Let $u,v,w$ be $f(a,g(h(a,a),c),b)$, $f(a,g(h(b,b),c),b)$ and $g(f(a,b,c),b)$, respectively. Then, $\mathtt{JointCon}(u,v,0) = \mathtt{JointCon}(u,w,0) = [\cdot]$, $\mathtt{JointCon}(u,v,1) = f(a,[\cdot],b)$, $\mathtt{JointCon}(u,w,1)$ is undefined, $\mathtt{JointCon}(u,v,2) = f(a,g([\cdot],c),b)$, and $\mathtt{JointCon}(u,v,3)$ is undefined.

LEMMA 3.3. *Let $s,t,u,v$ be terms with $u \neq v$. Let $\sigma$ be a solution of $\{F(s) \doteq u, F(t) \doteq v\}$. Then $\sigma(F) = \mathtt{JointCon}(u,v,|\mathtt{hp}(\sigma(F))|)$.*

PROOF. We prove the claim by induction on $|\mathtt{hp}(\sigma(F))|$. If $|\mathtt{hp}(\sigma(F))|$ is 0, then $\sigma(F)$ is $[\cdot]$, which coincides with $\mathtt{JointCon}(u,v,|\mathtt{hp}(\sigma(F))|)$. Now, suppose that $|\mathtt{hp}(\sigma(F))|$ is $l+1$ for some natural number $l$. This implies that $\sigma(F)$ is of the form $f(w_1,\ldots,w_{i-1},C[\cdot],w_{i+1},\ldots,w_m)$ for some function symbol $f$ and some $i \in [1,m]$. Since $\sigma$ is a solution of $\{F(s) \doteq u, F(t) \doteq v\}$, then $u$ and $v$ are of the form $f(u_1,\ldots,u_m)$ and $f(v_1,\ldots,v_m)$, respectively. For the same reason, $w_j = u_j = v_j$ for all $j \neq i$, and moreover, $\sigma(C[s]) = u_i$ and $\sigma(C[t]) = v_i$. Since $u \neq v$ holds, we also have $u_i \neq v_i$. Consider a new context variable $F'$ and the extension of $\sigma$ as $\sigma(F') = C[\cdot]$. Then, $\sigma$ is also a solution of $\{F'(s) \doteq u_i, F'(t) \doteq v_i\}$. Note that $|\mathtt{hp}(\sigma(F'))|$ is $l$, which is smaller than $|\mathtt{hp}(\sigma(F))|$. By induction hypothesis, $\sigma(F') = \mathtt{JointCon}(u_i,v_i,|\mathtt{hp}(\sigma(F'))|)$. Hence, we conclude $\sigma(F) = f(w_1,\ldots,w_{i-1},C[\cdot],w_{i+1},\ldots,w_m) = f(w_1,\ldots,w_{i-1},\sigma(F'),w_{i+1},\ldots,w_m) = f(w_1,\ldots,w_{i-1},\mathtt{JointCon}(u_i,v_i,|\mathtt{hp}(\sigma(F'))|),w_{i+1},\ldots,w_m) = \mathtt{JointCon}(u,v,|\mathtt{hp}(\sigma(F))|)$ □

### 3.3 The Intuition Behind the Algorithm

The algorithm is presented as a set of non-deterministic rules, since this is easier to explain. When we reason about its complexity, we argue about the determinized version that computes all guessing possibilities.

As already mentioned, we cannot directly guess a context of the right-hand side for every context variable, since there may be exponentially many contexts. In spite of this fact, we show that making an adequate use of the cases where the joint context can be inferred, the number of possibilities for each guessing can be drastically reduced. This fact allows us to use this approach also for the case when terms are represented with dags.

After some standard applications of simplification and first-order variable elimination, we can assume that any match-equation in the set $\Delta$ is of the form $F(s) \doteq t$, for some context variable $F$. Now, our goal is to remove one context variable by performing a guess, where the overall number of possibilities remains polynomial.

Suppose first that $\Delta$ contains two equations of the form $F(s_1) \doteq t_1$ and $F(s_2) \doteq t_2$ with $t_1 \neq t_2$. Then we can infer the context as in the last subsection. However, we still need the length of the hole position of $\sigma(F)$, for a possible solution $\sigma$. But this length can be guessed from $[0, \mathtt{min}(\mathtt{height}(t_1), \mathtt{height}(t_2))]$ which is linear in the input size.

Another situation is when $\Delta$ is of the form $\{F(s_1) \doteq t, \ldots, F(s_n) \doteq t\} \cup \Delta'$ for some term $t$ and $F$ does not occur elsewhere. In this case, a solution $\sigma$ for $\Delta$ necessarily satisfies that $\sigma(F)$ is a certain context $C$ such that $t$ is of the form $C[t']$ for some subterm $t'$ of $t$. Although there are exponentially many ways of choosing $C$, any of them can be used. Hence, we only have to look for $t'$, which can be guessed among only a linear number of possibilities. Then the problem can be reduced to $\{s_1 \doteq t', \ldots, s_n \doteq t'\} \cup \Delta'$. Note that the variable $F$ does not appear any more.

Now, suppose that some context variable has an occurrence at some non-root position in some term occurring in $\Delta$. A particular case occurs when there is an equation $F(s) \doteq t$ in $\Delta$ such that a subterm of $s$ is of the form $F(s')$, i.e. the context variable $F$ appears twice, at the root, and at some other position. Any possible solution $\sigma$ satisfies that either $\sigma(F)$ is the empty context $[\cdot]$, which can be decided with a guessing, or else $\sigma(F(s'))$ equals a proper subterm $t'$ of $t$. In the latter case, the pair of equations $\{F(s) \doteq t, F(s') \doteq t'\}$ with $t \neq t'$ allows us to proceed again by inferring the context, as in the first case.

If none of the previous cases hold, then there exist equations $F_1(s_1) \doteq t_1, F_2(s_2) \doteq t_2, \ldots, F_n(s_n) \doteq t_n$ in $\Delta$, where $F_1$ occurs in $s_2$, $F_2$ occurs in $s_3$, and so on, and $F_n$ occurs in $s_1$. In this sequence there is a maximal height term, say $t_1$. Thus, $\mathtt{height}(t_1) \geq \mathtt{height}(t_2)$. Note that $s_2$ contains a subterm of the form $F_1(s'_2)$. Then, similarly as above, either $\sigma(F_2) = [\cdot]$ or we can use the equations $F_1(s_1) \doteq t_1, F_1(s'_2) \doteq t'_2$, with $t'_2$ chosen from the proper subdags of $t_2$, to infer $\sigma(F_2)$.

With this approach each one of the $k$ context variables is instantiated by a guessing among a polynomial number of possibilities. Hence, at this point we can bring forward that the final cost of the algorithm will be exponential in $k$, which is a constant of the problem. However, we also need to choose a representation for dags that allows to efficiently instantiate both first-order and context variables. This is done in the next section.

### 3.4 Dag Representation of the k-CMD Algorithm

Before presenting our algorithm for the k-CMD problem in detail, it is necessary to define how we represent dags and how our algorithm deals with such a representation. As stated in Definition 2.5, dags can be represented as a DAG, which is a particular case of an STG, i.e an STG which does not have context non-terminals. For reasons that will be made clear soon we encode dags using this representation.

*Definition* 3.4. An instance of the k-context-matching problem with dags is a pair $\langle \Delta, G \rangle$ where the STG $G$ is a dag and $\Delta$ is a set of equations $\{A_{s_1} \doteq A_{t_1}, \ldots, A_{s_n} \doteq A_{t_n}\}$, where each $A_{s_i}$ and each $A_{t_i}$ is a term non-terminal of $G$, and each $w_{A_{t_i}}$ is ground. The question is to compute a substitution $\sigma$ (the solution) for the variables such that $\sigma(w_{A_{s_i}}) = w_{A_{t_i}}$ for every equation $i$ in $\Delta$.

During the execution of the k-CMD algorithm, the equations are processed, and $G$ is transformed in order to represent the partial solution at each step. More concretely, first-order variables are converted into term non-terminals, and context variables are converted into context non-terminals, whose generated terms and contexts represent substitutions of a partial solution. By *variables* we mean the variables of the problem and by *function symbols* we mean the terminals of the grammar which are not variables although, initially, all of them are terminals of the grammar. The initial $G$ has no context non-terminals, and it may incorporate them in order to represent that the context variables have been instantiated.

Our algorithm just instantiates variables by transforming them into non-terminals and adding rules for them, but does not change original rules. This ensures that right-hand sides of equations always represent subterms of an original $w_{G,A_{t_i}}$. Hence, although context variables are created during the execution, right-hand sides are always represented by a subset of the initial $G$, which continues being a dag according to Definition 2.5.

Using STGs for describing dags, instead of just talking about dags understood as directed acyclic graphs, has several advantages. First, we do not have to think about nodes and arrows. STGs are more syntactic and it is easier and clearer to add or remove rules to/from an STG than to talk about redirecting arrows, new inserted nodes, etc. Second, the formalism of STGs is an improvement in clarity and simplicity with respect to the usual concept of solved form for representing partial and final solutions. At the end of the execution, the obtained substitution for a first order variable $x$ will be $w_x$, i.e. this variable will be a term non-terminal, and its generated term will be the substitution computed for it. Analogously, a context variable $F$ will be transformed into a context non-terminal, and the substitution computed for it will be $w_F$. Third, analyzing the size increase of the representation due to variable instantiation is much simpler: adding a rule $F \to \alpha$ for a context variable $F$ and transforming $F$ into a context non-terminal is easy to analyze, whereas replacing each node in the dag labeled with $F$ by new nodes representing its substitution is a more complicated operation. On the other hand, this representation has the disadvantage that the set of equations is not enough by itself, but needs the STG. For this reason, our algorithm needs to use the rules of $G$ and perform some replacements of non-terminals by their corresponding definition.

There is a case where our algorithm has to guess a partial solution from an exponential number of possibilities. This happens when we have equations $F(s_1) \doteq t, \ldots, F(s_n) \doteq t$, and the context variable $F$ does not appear elsewhere. In this case, the only important information to be kept is which subterm $t'$ of $t$ has to be selected in order to generate the equations $s_1 \doteq t', \ldots, s_n \doteq t'$. The solution for $F$ might be any context $C$ such that $Ct' = t$, that is, the hole position of the solution of $F$ is any path from the root of $t$ to an occurrence of $t'$. We do not want to guess $C$ from an exponential number of possibilities. Unfortunately, these exponentially large set of contexts cannot be represented by $G$. For this reason, in the algorithm we have a third component, apart from the set of equations $\Delta$ and the STG $G$, representing the possible elections for the variables $F$ of this kind. This component is a set of expressions of the form $F \in \mathtt{Contexts}(A, A')$, representing that $F$ can be replaced by any context $C$ such that $Cw_{A'} = w_A$. Hence, our algorithm deals with triples $\langle \Delta, G, \Gamma \rangle$ where $\Gamma$ is the set containing this kind of expressions.

As a last ingredient, we need to adapt the operation $\mathtt{JointCon}$, presented in Section 3.2, to our representation.

*Definition* 3.5. Let $G$ be an $STG$ and $A, B$ be two term non-terminals of $G$ such that $w_A \neq w_B$ and $\mathtt{restriction}(G, \{A, B\})$ is a DAG representing ground terms. Let $l$ be a natural number.

Then, $\mathtt{JointCG}(G, A, B, l)$ is defined as an extension of $G$ recursively as follows. $\mathtt{JointCG}(G, A, B, 0)$ contains $G$ plus the rule $C \to [\cdot]$, where $C$ is a new context non-terminal. For the case of $\mathtt{JointCG}(G, A, B, l+1)$, if the rules of $A$ and $B$ are of the form $A \to f(A_1, \ldots, A_{i-1}, A_i, A_{i+1}, \ldots, A_n)$ and $B \to f(A_1, \ldots, A_{i-1}, B_i, A_{i+1}, \ldots, A_n)$, for some $i$ satisfying $w_{A_i} \neq w_{B_i}$, then $\mathtt{JointCG}(G, A, B, l+1)$ contains $\mathtt{JointCG}(G, A_i, B_i, l)$, which has a context non-terminal $C'$ generating $\mathtt{JointCon}(w_{A_i}, w_{B_i}, l)$, plus the rule $C \to f(A_1, \ldots, A_{i-1}, C', A_{i+1}, \ldots, A_n)$, where $C$ is a new context non-terminal. In any other case, $\mathtt{JointCG}(G, A, B, l+1)$ is undefined.

LEMMA 3.6. *Let $G$ be an STG and $A, B$ be two term non-terminals of $G$ such that $w_A \neq w_B$ and $\mathtt{restriction}(G, \{A, B\})$ is a DAG representing ground terms. Let $l$ be a natural number. Assume also that $\mathtt{restriction}(G, \{A, B\})$ is compressed optimally, i.e. equal terms are represented by the same term non-terminal.*

*Then, $\mathtt{JointCG}(G, A, B, l)$ adds at most $\mathtt{depth}(G)$ new context non-terminals to $G$, and has one symbol generating $\mathtt{JointCon}(w_A, w_B, l)$. Moreover, all the added context non-terminals $C$ have rules which are of the form $C \to [\cdot]$ or $C \to f(A_1, \ldots, A_{i-1}, C', A_{i+1}, \ldots, A_n)$, where the terminal $f$ is necessarily a function symbol, i.e. it is not a variable. The time complexity of this construction is $\mathcal{O}(\mathtt{depth}(G))$.*

*Definition* 3.7. Let $G$ be an $STG$ and $A, B$ be two term non-terminals of $G$ such that $w_A \neq w_B$ and $\mathtt{restriction}(G, \{A, B\})$ is a DAG representing ground terms.. Let $F$ be a context variable which is a terminal of arity 1 of $G$. Let $l$ be a natural number. Assume also that $\mathtt{restriction}(G, \{A, B\})$ is compressed optimally, i.e. equal terms are represented by the same term non-terminal.

Then, $\mathtt{JointCGF}(G, F, A, B, l)$ is an STG obtained from $\mathtt{JointCG}(G, A, B, l)$, which has a context non-terminal $C$ not occurring in $G$ and generating the context $\mathtt{JointCon}(w_A, w_B, l)$, by transforming $F$ into a context non-terminal, and replacing the non-terminal $C$ by $F$ everywhere.

### 3.5 Rules of the k-CMD Algorithm

*Definition* 3.8. The $k$-CMD algorithm is presented in figures 1, 2 and 3 as a set of transformation rules which deal with triples $\langle \Delta, G, \Gamma \rangle$, where $\Delta$ is a set of equations defined over an STG $G$, where the right hand sides of equations are non-terminals in a DAG representing ground terms, and $\Gamma$ is a set of expressions each one representing all solutions for a context variable, as described in the previous section. We assume that, initially, equal subterms in the right-hand sides of equations are represented by the same term non-terminal, i.e. optimal dag compression is used.

$$
\begin{array}{ll}
\text{UNFOLD1:} & \dfrac{\langle \Delta \cup \{A \doteq B\}, G = (\mathcal{TN} \cup \{A, B\}, \mathcal{CN}, \Sigma, R \cup \{A \to u\}), \Gamma \rangle}{\langle \Delta \cup \{u \doteq B\}, G, \Gamma \rangle} \\[2ex]
\text{UNFOLD2:} & \dfrac{\begin{array}{l}\langle \Delta \cup \{CA \doteq B\}, G = (\mathcal{TN} \cup \{A, B\}, \mathcal{CN} \cup \{C\}, \Sigma, \\ R \cup \{C \to f(A_1, \ldots, C_i, \ldots, A_m)\}), \Gamma \rangle \end{array}}{\langle \Delta \cup \{f(A_1, \ldots, C_iA, \ldots, A_m) \doteq B\}, G, \Gamma \rangle} \\[2ex]
\text{UNFOLD3:} & \dfrac{\langle \Delta \cup \{CA \doteq B\}, G = (\mathcal{TN} \cup \{A, B\}, \mathcal{CN} \cup \{C\}, \Sigma, R \cup \{C \to [\cdot]\}), \Gamma \rangle}{\langle \Delta \cup A \doteq B\}, G, \Gamma \rangle}
\end{array}
$$

Fig. 1. Unfold-Rules of the $k$-CMD Algorithm

$$
\text{DECOMPOSE:} \quad \dfrac{\begin{array}{l}\langle \Delta \cup \{f(u_1, \ldots, u_m) \doteq B\}, G = (\mathcal{TN} \cup \{B_1, \ldots, B_m\}, \\ \mathcal{CN}, \Sigma \cup \{f\}, R \cup \{B \to f(B_1, \ldots, B_m)\}), \Gamma \rangle \end{array}}{\langle \Delta \cup \{u_1 \doteq B_1, \ldots, u_m \doteq B_m\}, G, \Gamma \rangle}
$$

where $f$ is a function symbol ($m = \mathtt{arity}(f)$)

$$
\text{FAIL:} \quad \dfrac{\begin{array}{l}\langle \Delta \cup \{f(u_1, \ldots, u_m) \doteq B\}, G = (\mathcal{TN} \cup \{B_1, \ldots, B_{m'}\}, \\ \mathcal{CN}, \Sigma \cup \{f, g\}, R \cup \{B \to g(B_1, \ldots, B_{m'})\}), \Gamma \rangle \end{array}}{\bot}
$$

where $f \neq g$ are function symbols ($m = \mathtt{arity}(f)$, $m' = \mathtt{arity}(g)$)).

$$
\text{ELIMX:} \quad \dfrac{\langle \Delta \cup \{x \doteq B\}, G = (\mathcal{TN} \cup \{B\}, \mathcal{CN}, \Sigma \cup \{x\}, R), \Gamma \rangle}{\langle \Delta \cup \{x \doteq B\}, G' = (\mathcal{TN} \cup \{B, x\}, \mathcal{CN}, \Sigma, R \cup \{x \to B\}), \Gamma \rangle}
$$

where $x$ is a first-order variable and a terminal.

Fig. 2. First-Order-Rules of the $k$-CMD Algorithm

This will hold during the execution. Given an instance of the problem $\langle \{A_{s_1} \doteq A_{t_1}, \ldots, A_{s_n} \doteq A_{t_n}\}, G \rangle$, the starting triple is $\langle \{A_{s_1} \doteq A_{t_1}, \ldots, A_{s_n} \doteq A_{t_n}\}, G, \emptyset \rangle$, and the constant $L$ occurring in the rules is $\max_{1 \leq i \leq n}(\mathtt{height}(w_{G, A_{t_i}}))$.

There are two kinds of choices the algorithm can do. On the one side there are the "don't care" selections, which include the strategy stating which rule is applied and the selection of the equations involved in the rule application. On the other side we have the guessings, which make the algorithm non-deterministic. Those correspond to the decisions marked as "guessed" in the conditions of the rules, but also to the selection performed when the resulting part of a rule has a disjunction.

We differentiate our set of inference rules in two disjoint subsets. We call the first rules *unfolding* rules (see Fig. 1), since their purpose is to replace the non-terminals of $G$ occurring in the equations by their definition in $G$. Hence, these rules are related to our grammar-based representation for dags. We refer to the rest of the rules as *solving* rules, since they represent the actual algorithm as described in Section 3.3; these are splitted into the first-order rules (see Fig. 2) and the context-variable rules (see Fig. 3). The application of *solving* rules transforms the set of equations into a new set. Depending on the case, more than one rule can be applied to a given set of equations. Hence, the inference system represents, in fact, a family of algorithms, depending on the strategy for deciding which rule to apply and to which subset of equations. As commented before, our initial set of equations is of the form $\{A_{s_1} \doteq A_{t_1}, \ldots, A_{s_n} \doteq A_{t_n}\}$. But after applying the transformation rules, the form of these equations may change. Nevertheless, at any step of the algorithm the current equations are *simple*, according to the following definition.

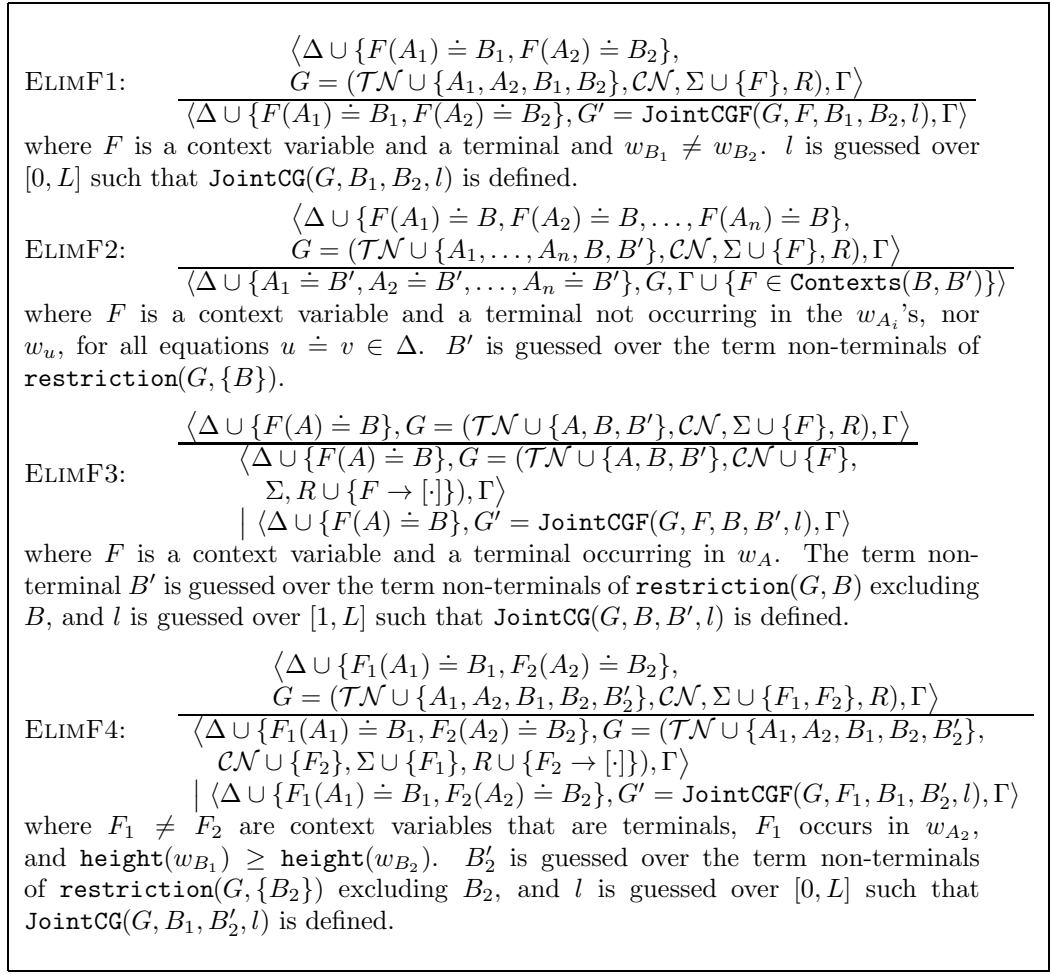

Fig. 3. Elim-F-Rules of the $k$-CMD Algorithm

*Definition* 3.9. Let $G = (\mathcal{TN}, \mathcal{TC}, \Sigma, R)$ be an STG, and let $u \doteq v$ be an equation, where $u, v \in \mathcal{T}(\mathcal{TN} \cup \mathcal{TC} \cup \Sigma)$. The equation $u \doteq v$ is called *simple* over $G$ if it is of one of the following forms.

— $A \doteq B$
— $CA \doteq B$
— $\alpha(A_1, \ldots, A_m) \doteq B$

—$f(A_1, \ldots, A_{i-1}, C_i A, A_{i+1}, \ldots, A_m) \doteq B$,

where $A$ is a term non-terminal of $G$, $B$ is a term non-terminal of $G$ representing a ground term, $C$ is a context non-terminal of $G$, and the terms $\alpha(A_1, \ldots, A_m)$ and $f(A_1, \ldots, A_{i-1}, C_i, A_{i+1}, \ldots, A_m)$ are right-hand sides of rules of $G$, for a terminal $\alpha$, a terminals $f$, which is also function symbols, term non-terminals $A_1, \ldots, A_m$, and a context non-terminal $C_i$. Variables can only occur as some $\alpha$.

The following lemma shows that no rule of the form $C \to C_1 C_2$ occurs in the $k$-CMD algorithm.

LEMMA 3.10. *Let $\langle \Delta, G, \Gamma \rangle$ be a triple obtained by our algorithm at any point of the execution. Then, the rules of $G$ are of the following forms.*

—$A \to A_1$
—$A \to \alpha(A_1, \ldots, A_m)$,
—$A \to C A_1$
—$C \to f(A_1, \ldots, A_{i-1}, C_i, A_{i+1}, \ldots, A_m)$
—$C \to [\cdot]$

*where $A, A_1, A_2, \ldots, A_m$ are term non-terminals of $G$, $C, C_i$ are context non-terminals of $G$, $\alpha$ is a terminal of $G$, and $f$ is a terminal of $G$, which is also a function symbol.*

PROOF. We prove the lemma by induction on the number of applied inference rules. For the base case, note that the lemma holds for the STG $G_0$ given as input since, by Definitions 2.5 and 3.4, all the rules in $G_0$ are of the form $A \to \alpha(A_1, \ldots, A_m)$ for some term non-terminals $A_1, \ldots, A_m$ and a terminal $\alpha$ of $G_0$.

For the induction case, let $\langle \Delta', G', \Gamma' \rangle$ be the triple from which $\langle \Delta, G, \Gamma \rangle$ was obtained by an inference rule application. By induction hypothesis $\langle \Delta', G', \Gamma' \rangle$ satisfies the conditions of the lemma. We distinguish cases according to the inference rule applied to $\langle \Delta', G', \Gamma' \rangle$ in order to show that the rules in $G$ follow the conditions of the lemma. Note that for the inference rules that do not modify the STG (*unfolding* rules, DECOMPOSE, FAIL, and ELIMF2), this is straightforward. Otherwise, if ELIMX was the applied rule, $x$ became a term non-terminal and a rule of the form $x \to A$ was added to $G'$ for some terminal $x$ representing a first-order variable and term non-terminal $A$. Note that the added rule satisfies the conditions of the lemma. Finally, if the applied rule was either ELIMF1, ELIMF3, or ELIMF4 then either $G$ was extended by the `JointCon` construction or a rule $F \to [\cdot]$ was added to $G'$, for some context variable $F$. By Lemma 3.6, in both cases all the added rules satisfy the condition of the lemma. □

LEMMA 3.11. *Let $\langle \Delta, G, \Gamma \rangle$ be the triple obtained by our algorithm at a point of the execution. Then, the set $\Delta$ consists of* simple *equations over $G$.*

PROOF. Since for the triple given as input $\langle \Delta_0, G_0, \Gamma_0 = \emptyset \rangle$ all the equations in $\Delta$ are of the form $A_s \doteq A_t$ for some term non-terminals $A_s, A_t$ in $G_0$, the statement of the Lemma holds in this case. Hence, for proving this lemma it suffices to check that after an inference step where $\langle \Delta, G, \Gamma \rangle$ was obtained from a triple $\langle \Delta', G', \Gamma' \rangle$, each new equation in $\Delta$ is *simple* over $G$. Checking this is an easy task for rules FAIL, ELIMX, ELIMF1, ELIMF2, ELIMF3, ELIMF4, UNFOLD2, and UNFOLD3, since the new produced equations are explicitly defined. For DECOMPOSE the result follows by induction hypothesis. Finally, the produced equations due to the application of UNFOLD1 are of the form $u \doteq B$, where $B$ is a term non-terminal and $u$ corresponds to a right-hand side of a rule in $G'$ and hence, they satisfy the condition to be simple over $G$ due to Lemma 3.10. □

Before proving soundness, completeness and termination of our inference system we should define a notion of solution of the triples the k-CMD algorithm deals with.

*Definition* 3.12. A solution of $\langle \Delta, G, \Gamma \rangle$ is a substitution $\sigma$ such that $\sigma(w_{G,u}) = w_{G,v}$ for each equation $u \doteq v$ in $\Delta$, $\sigma(w_{G,x}) = \sigma(x)$ for each first-order variable $x$, and $\sigma(w_{G,F}) = \sigma(F)$ for each context variable $F$. Furthermore, for each context variable $F$ such that $(F \in \texttt{Contexts}(A_1, A_2)) \in \Gamma$, where $A_1$ and $A_2$ are term non-terminals of $G$, it holds that $\sigma(F)w_{G,A_2} = w_{G,A_1}$.

Let $\langle \Delta, G, \Gamma \rangle$ be a triple generated by our algorithm at any point of the execution. Note that some of the variables may have been isolated and, hence, the STG $G$ was extended in order to represent the corresponding instantiations. As stated in the previous definition, a solution of $\langle \Delta, G, \Gamma \rangle$ has to be consistent with this extensions. The following lemma, together with the definition of a solution $\sigma$ of $\langle \Delta, G, \Gamma \rangle$, states that our representation for partial solutions by extending the grammar is correct in the sense that the same term is obtained by applying a solution to the term generated by $G$ before and after such an extension. It will be helpful when proving soundness and completeness.

LEMMA 3.13. *Let $G = (\mathcal{TN}, \mathcal{CN}, \Sigma \cup \{V\}, R)$ be an STG obtained at any point of the execution of the k-CMD algorithm. Let $V$ be a terminal of $G$ representing either a first-order or a context variable. Let $G'$ be the STG obtained from $G$ by converting $V$ into a non-terminal of the grammar and adding some new rules and non-terminals such that $V$ generates a certain term or context $w_{G',V}$. Let $\sigma$ be a substitution such that $\sigma(V) = \sigma(w_{G',V})$. Let $t$ be a term in $\mathcal{T}(\mathcal{TN} \cup \mathcal{CN} \cup \Sigma \cup \{V\})$ or a context in $\mathcal{C}(\mathcal{TN} \cup \mathcal{CN} \cup \Sigma \cup \{V\})$. Then, $\sigma(w_{G,t}) = \sigma(w_{G',t})$.*

PROOF. The proof is an easy induction on the size of $t$ and the number of rule application s to derive $w_{G,t}$. □

LEMMA 3.14. *The set of rules is sound.*

PROOF. Let $\langle \Delta', G', \Gamma' \rangle$ be the triple obtained by our algorithm by applying an inference step on $\langle \Delta, G, \Gamma \rangle$. By inspecting the rules, we can check that every solution $\sigma$ of $\langle \Delta', G', \Gamma' \rangle$ is also a solution of $\langle \Delta, G, \Gamma \rangle$: We distinguish cases depending on which rule was applied for obtaining $\langle \Delta', G', \Gamma' \rangle$ from $\langle \Delta, G, \Gamma \rangle$.

Note that the rules ELIMX, ELIMF1, ELIMF3 and ELIMF4 instantiate either a first-order or a context variable $V$. Therefore, if one of those rules was the rule applied to $\langle \Delta, G, \Gamma \rangle$ then $G'$ was obtained from $G$ by transforming $V$ into a non-terminal of the STG and adding some non-terminals and their corresponding rules such that $V$ generates $w_{G',V}$. By Definition 3.12, for being a solution of $\langle \Delta', G', \Gamma' \rangle$, $\sigma$ satisfies $\sigma(V) = \sigma(w_{G',V})$. Hence, $G$ and $G'$ satisfy the conditions of Lemma 3.13 and we can conclude $\sigma(w_{G,t}) = \sigma(w_{G',t})$ for every term $t$ in $\mathcal{T}(\mathcal{TN} \cup \mathcal{CN} \cup \Sigma)$, where $G = (\mathcal{TN}, \mathcal{CN}, \Sigma, R)$. It follows that $\sigma(x) = \sigma(w_{G,x})$ for every first-order variable $x$, and $\sigma(F) = \sigma(w_{G,F})$ for every context variable $F$. Moreover, since none of these rules changed neither the set $\Delta$ nor $\Gamma$, $\sigma$ is also a solution for $\langle \Delta, G, \Gamma \rangle$.

Suppose the rule applied is ELIMF2. In this case, $G' = G$ but both sets $\Delta$ and $\Gamma$ are changed. Concretely, a set of equations of the form $\{F(A_1) \doteq B, F(A_2) \doteq B, \ldots, F(A_n) \doteq B\}$ of $\Delta$ is replaced by a set of equations of the form $\{A_1 \doteq B', A_2 \doteq B', \ldots, A_n \doteq B'\}$ to obtain $\Delta'$ and the restriction $F \in \texttt{Contexts}(B, B')$ was added to $\Gamma$ to obtain $\Gamma'$. By Definition 3.12, since $\sigma$ is a solution of $\langle \Delta', G', \Gamma' \rangle$, it holds $\sigma(w_{G',A_i}) = w_{G',B'}$ for each $i \in [1, n]$, and $\sigma(F)w_{G',B'} = w_{G',B}$. Since $G = G'$ and $\Delta - \{F(A_i) \doteq B \mid i \in [1, n]\} = \Delta' - \{A_i \doteq B' \mid i \in [1, n]\}$, it suffices to prove that $\sigma(w_{G,F(A_i)}) = w_{G,B}$ for $i \in [1, n]$ to show that $\sigma$ is also a solution of $\langle \Delta, G, \Gamma \rangle$. Since $G = G'$ and $\sigma(w_{G',A_i}) = w_{G',B'}$ then $\sigma(w_{G,A_i}) = w_{G,B'}$ holds. Furthermore, it holds that $\sigma(w_{G,F(A_i)}) = \sigma(F(w_{G,A_i})) = \sigma(F)\sigma(w_{G,A_i}) = \sigma(F)\sigma(w_{G,B'}) = \sigma(F)\sigma(w_{G',B'}) = w_{G',B} = w_{G,B}$. Hence, we proved that $\sigma(w_{G,F(A_i)}) = w_{G,B}$ and thus $\sigma$ is also a solution of $\langle \Delta, G, \Gamma \rangle$.

For rule FAIL, it is obvious that the assumption of a solution $\sigma$ for the resulting triple $\langle \Delta', G', \Gamma' \rangle$ cannot be satisfied.

Suppose the rule applied is DECOMPOSE. Then, $G' = G$, $\Gamma = \Gamma'$ and an equation $f(u_1, \ldots, u_m) \doteq B$ in $\Delta$ where $B \to f(B_1, \ldots, B_m)$ is the rule in $G$ is replaced by the equations $u_1 \doteq B_1, \ldots, u_m \doteq B_m$ to obtain $\Delta'$. Hence, it suffices to prove that $\sigma(w_{G,f(u_1,\ldots,u_m)}) = w_{G,f(B_1,\ldots,B_m)}$ in order to show that $\sigma$ is also a solution for $\langle \Delta, G, \Gamma \rangle$. Since $\sigma$ is a solution of $\langle \Delta', G', \Gamma' \rangle$ it holds $\sigma(w_{G',u_1}) = w_{G',B_1}, \ldots, \sigma(w_{G',u_m}) = w_{G',B_m}$. Thus, $\sigma(w_{G,f(u_1,\ldots,u_m)}) = \sigma(w_{G',f(u_1,\ldots,u_m)}) = f(\sigma(w_{G',u_1}), \ldots, \sigma(w_{G',u_m})) = f(w_{G',B_1}, \ldots, w_{G',B_m}) = f(w_{G,B_1}, \ldots, w_{G,B_m}) = w_{G,f(B_1,\ldots,B_m)}$.

In the case where the rule applied is an *unfolding rule*, note that these rules just replace non-terminals of $G$ by their definition in $G$. Hence, since $w_N = w_\alpha$ for each non-terminal $N$ with a rule $N \to \alpha \in G$, every solution of $\langle \Delta', G', \Gamma' \rangle$ is also a solution of $\langle \Delta, G, \Gamma \rangle$. □

The following lemma is an adaptation of Lemma 3.3 to our STG-based representation for dags, which will be helpful when proving completeness.

LEMMA 3.15. *Let $G = (\mathcal{TN}, \mathcal{CN}, \Sigma, R)$ be an STG. Let $u_1, u_2$ be terms in $\mathcal{T}(\mathcal{TN} \cup \mathcal{CN} \cup \Sigma)$. Let $A_1, A_2, B_1$ and $B_2$ be term non-terminals of $G$ such that $w_{G,B_1} \neq w_{G,B_2}$ and both $w_{G,B_1}$ and $w_{G,B_2}$ are ground. Let $\texttt{restriction}(G, B_1, B_2)$ be compressed optimally as a DAG. Let $\sigma$ be a solution of $\langle \{F(u_1) \doteq B_1, F(u_2) \doteq B_2\}, G, \Gamma = \emptyset \rangle$ where the context variable $F$ is a terminal of $G$. Let $G' = \texttt{JointCGF}(G, F, B_1, B_2, |\texttt{hp}(\sigma(F))|)$. Then, $\sigma(F) = w_{G',F}$.*

PROOF. This lemma directly follows from Lemma 3.3 and Definition 3.7. □

LEMMA 3.16. *Every rule is complete. I.e. for every solution $\sigma$ of $\langle \Delta, G, \Gamma \rangle$, and for every rule application, there is a result $\langle \Delta', G', \Gamma' \rangle$ such that $\sigma$ is also a solution of $\langle \Delta', G', \Gamma' \rangle$. Moreover, any maximal sequence of rule applications computes a representation of all solutions, by gathering all guesses and alternatives in the rules.*

PROOF. Let $\sigma$ be a solution for some triple $\langle \Delta, G, \Gamma \rangle$ obtained by our algorithm. It suffices to show that after applying any applicable rule to $\langle \Delta, G, \Gamma \rangle$, one of the resulting triples $\langle \Delta', G', \Gamma' \rangle$ among the possible guesses also has $\sigma$ as solution. We distinguish cases depending on which inference step was applied for obtaining $\langle \Delta', G', \Gamma' \rangle$ from $\langle \Delta, G, \Gamma \rangle$. We state explicitly here $G = (\mathcal{TN}, \mathcal{CN}, \Sigma, R)$ because it will be necessary, in some cases, to refer to the set of terms $\mathcal{T}(\mathcal{TN} \cup \mathcal{CN} \cup \Sigma)$.

Assume the applied rule is DECOMPOSE. Then, $G' = G$, $\Gamma = \Gamma'$ and an equation $f(u_1, \ldots, u_m) \doteq B$ in $\Delta$ with rule $B \to f(B_1, \ldots, B_m)$ is replaced by the equations $u_1 \doteq B_1, \ldots, u_m \doteq B_m$ to obtain $\Delta'$, where each $u_i \in \mathcal{T}(\mathcal{TN} \cup \mathcal{CN} \cup \Sigma)$. Hence, it suffices to prove $\sigma(w_{G',u_1}) = w_{G',B_1}, \ldots, \sigma(w_{G',u_m}) = w_{G',B_m}$ in order to show that $\sigma$ is also a solution for $\langle \Delta', G', \Gamma' \rangle$. Since $\sigma$ is a solution of $\langle \Delta, G, \Gamma \rangle$, it holds that $\sigma(w_{G,f(u_1,\ldots,u_m)}) = w_{G,f(B_1,\ldots,B_m)}$ which implies $\sigma(f(w_{G,u_1}, \ldots, w_{G,u_m})) = f(w_{G,B_1}, \ldots, w_{G,B_m})$, and hence $\sigma(w_{G,u_1}) = w_{G,B_1}, \ldots, \sigma(w_{G,u_m}) = w_{G,B_m}$. Finally, since $G = G'$, $\sigma$ is also a solution of $\langle \Delta', G', \Gamma' \rangle$.

Assume the applied rule is ELIMX. Then $\Gamma = \Gamma'$ and $\Delta = \Delta'$. For a concrete equation $x \doteq B \in \Delta$, $G$ was extended to $G'$ by converting $x$ into a term non-terminal and adding the rule $x \to B$. Since $\sigma$ is a solution of $\langle \Delta, G, \Gamma \rangle$ and $x$ is a terminal of $G$, $w_{G,x} = x$ and $\sigma(x) = w_{G,B}$ holds. Furthermore, $w_{G',x} = w_{G',B} = w_{G,B}$ since $B$ is the definition of $x$ in $G'$ and none of the rules of $G$ were changed to obtain $G'$. Hence, $\sigma(x) = w_{G,B} = w_{G',B} = w_{G',x} = \sigma(w_{G',x})$, where the last equality holds because $w_{G',x}$ is ground. Thus, we can apply Lemma 3.13 and claim that, for every term $t$ in $\mathcal{T}(\mathcal{TN} \cup \mathcal{CN} \cup \Sigma)$, $\sigma(w_{G,t}) = \sigma(w_{G',t})$. Hence, since $\Gamma = \Gamma'$ and $\Delta = \Delta'$, $\sigma$ is also a solution for $\langle \Delta', G', \Gamma' \rangle$.

For the FAIL rule it is clear that the assumption on the existence of a solution cannot be satisfied.

Suppose that the applied rule is ELIMF1. In this case, $\Delta = \Delta'$, $\Gamma = \Gamma'$ and $G$ was extended to $G'$ by converting the terminal $F$, which is a context variable, into a context non-terminal. Some rules and non-terminals were added such that $F$ generates a ground context $w_{G',F}$. We first show that $\sigma(F) = \sigma(w_{G',F})$ holds for one of the possible guesses when applying this rule.

Since $|\mathtt{hp}(\sigma(F))|$ is smaller than or equal to $L$ ($|\mathtt{hp}(\sigma(F))| \in [0, L]$) we can assume that $l$ is guessed as $|\mathtt{hp}(\sigma(F))|$ in the rule application. Then, by the conditions for this rule application, there are equations of the form $F(A_1) \doteq B_1, F(A_2) \doteq B_2$ in $\Delta$ such that $w_{B_1} \neq w_{B_2}$. Furthermore, both $w_{B_1}$ and $w_{B_2}$ are ground and $G'$ is constructed as $\mathtt{JointCGF}(G, F, B_1, B_2, |\mathtt{hp}(\sigma(F))|)$. Hence, by Lemma 3.15, $\sigma(F) = w_{G',F}$. Moreover, we can apply Lemma 3.13 and conclude that $\sigma(w_{G,t}) = \sigma(w_{G',t})$ for every term $t \in \mathcal{T}(\mathcal{TN} \cup \mathcal{CN} \cup \Sigma)$. Thus, $\sigma$ is a solution of $\langle \Delta', G', \Gamma' \rangle$.

Suppose now that the applied rule is ELIMF2. In this case, $G = G'$, and some equations of the form $F(A_1) \doteq B, F(A_2) \doteq B, \ldots, F(A_n) \doteq B$ of $\Delta$ such that $F$ does not occur in $w_u$ for any other equation $u \doteq v$ in $\Delta$ were replaced by the equations $A_1 \doteq B', \ldots, A_n \doteq B'$ to obtain $\Delta'$. Moreover, the restriction $F \in \mathtt{Contexts}(B, B')$ was added to $\Gamma$ to obtain $\Gamma'$. Since $\sigma$ is a solution of $\langle \Delta, G, \Gamma \rangle$, it is also a solution of $\{F(A_1) \doteq B, F(A_2) \doteq B, \ldots, F(A_n) \doteq B\}$. Hence, there exists some subterm $w_{G,B'}$ of $w_{G,B}$ satisfying $\sigma(w_{G,A_1}) = w_{G,B'}, \ldots, \sigma(w_{G,A_n}) = w_{G,B'}$ which corresponds to $w_{G,B}|_{\mathtt{hp}(\sigma(F))}$. In our representation choosing a subterm of $w_{G,B}$ is equivalent to choosing one of the term non-terminals of $\mathtt{restriction}(G, \{B\})$. Thus, we can consider the case where $B'$ is the term non-terminal guessed in the rule application. In this case $\sigma(w_{G',A_1}) = w_{G',B'}, \ldots, \sigma(w_{G',A_n}) = w_{G',B'}$ holds since $G = G'$. Therefore, $\sigma$ is also a solution for $\langle \Delta', G', \Gamma' \rangle$. With respect to $\sigma(F)$, it satisfies $\sigma(F) w_{G,B'} = w_{G,B}$, which is exactly the condition added to $\Gamma$ by the rule application in order to keep a representation of all possible instantiations for the context variable $F$.

Suppose that the applied rule is ELIMF3. In this case, $\Delta = \Delta'$, $\Gamma = \Gamma'$ and $G$ was extended to $G'$ by converting a terminal $F$ representing a context variable into a context non-terminal. Some rules and non-terminals were added such that $F$ generates the ground term $w_{G',F}$. We first show that $\sigma(F) = \sigma(w_{G',F})$ holds for one of the possible guesses when applying this rule.

By the condition of this rule application, $F(A) \doteq B$ is an equation in $\Delta$ where $F$ occurs in $w_A$. The case $\sigma(F) = [\cdot]$ is covered by the first alternative of the rule. Now assume that $\sigma(F) \neq [\cdot]$. Since $F$ occurs in $w_{G,A}$, there exists a proper subterm of $w_{G,F(A)}$ (a subterm of $w_{G,A}$) of the form $F(u)$ for some term $u \in \mathcal{T}(\Sigma)$. Since $\sigma(F(w_{G,A})) = w_{G,B}$ holds and $\sigma(F) \neq [\cdot]$, there exists a proper subterm $w_{G,B'}$ of $w_{G,B}$ such that $\sigma(F(u)) = w_{G,B'}$ and, for the same reason as in the previous case, $B'$ is a term non-terminal in $\texttt{restriction}(\texttt{G}, \texttt{B})$ excluding $B$. We consider the case where the term non-terminal $B'$ is guessed by the rule application and $l$ is guessed as $|\texttt{hp}(\sigma(F))|$. When these two guesses are done, $G'$ is constructed as $\texttt{JointCGF}(G, F, B, B', |\texttt{hp}(\sigma(F))|)$. Furthermore, we know that $\sigma$ satisfies $\sigma(F(u)) = w_{G,B'}$ and $\sigma(w_{G,F(A)}) = w_{G,B}$. Moreover, $w_{G,B'}$ and $w_{G,B}$ are ground, and $w_{B'} \neq w_B$, since $w_{B'}$ is a proper subdag of $w_B$. Hence, we can apply Lemma 3.15 and conclude $\sigma(F) = w_{G',F}$. As before, we can apply Lemma 3.13 and conclude that $\sigma$ is a solution of $\langle \Delta', G', \Gamma' \rangle$.

Suppose that the applied rule is ELIMF4. In this case, $\Delta = \Delta'$, $\Gamma = \Gamma'$ and $G$ was extended to $G'$ by either converting a terminal $F_2$ or a terminal $F_1 \neq F_2$, each of them representing a context variable, into a context non-terminal. Each of these cases corresponds to one of the two alternatives of the rule. In the first case the rule $F_2 \to [\cdot]$ was added, such that $F_2$ generates $w_{G',F_2} = [\cdot]$, the empty context. In the second case some rules and non-terminals were added, such that $F_1$ generates the ground context $w_{G',F_1}$. We first show that either $\sigma(F_2) = \sigma(w_{G',F_2})$, in the former case, or $\sigma(F_1) = \sigma(w_{G',F_1})$ in the latter case.

By the condition of the application of ELIMF4, there is a pair of equations in $\Delta$ of the form $F_1(A_1) \doteq B_1$ and $F_2(A_2) \doteq B_2$. Furthermore, $F_1$ occurs in $w_{G,A_2}$, and $\texttt{height}(w_{G,B_1}) \geq \texttt{height}(w_{G,B_2})$. The case $\sigma(F_2) = [\cdot]$ is covered by the first alternative of the rule, and it is obvious that $\sigma(F_2) = \sigma(w_{G',F_2}) = [\cdot]$ holds in this case. Now assume that $\sigma(F_2) \neq [\cdot]$. Since $F_1$ occurs in $w_{G,A_2}$, there exists a proper subterm of $w_{G,F_2(A_2)}$ (a subterm of $w_{G,A_2}$) of the form $F_1(u)$, for some $u \in \mathcal{T}(\mathcal{TN} \cup \mathcal{CN} \cup \Sigma)$. Moreover, since $\sigma(w_{G,F_2(A_2)}) = w_{G,B_2}$ holds, and $\sigma(F_2) \neq [\cdot]$, there exists a proper subterm $w_{G,B'_2}$ of $w_{G,B_2}$ such that $\sigma(F_1(u)) = w_{G,B'_2}$ and, for the same reason as in the previous case, $B'_2$ is represented by a term non-terminal in $\texttt{restriction}(\texttt{G}, \texttt{B}_2)$ excluding $B_2$, since the subterm is proper. We consider the case where the term non-terminal $B'_2$ is guessed by the rule application and $l$ is guessed as $|\texttt{hp}(\sigma(F_1))|$. Hence, $G'$ is constructed as $\texttt{JointCGF}(G, F_1, B_1, B'_2, |\texttt{hp}(\sigma(F_1))|)$. We know that $\sigma$ has to satisfy $\sigma(w_{F_1(A_1)}) = w_{G,B_1}$ and $\sigma(F_1(u)) = w_{G,B'_2}$. Moreover, $w_{G,B_1}$ and $w_{G,B'_2}$ are ground, and $w_{G,B_1} \neq w_{G,B'_2}$ holds, since $\texttt{height}(w_{G,B_1}) \geq \texttt{height}(w_{G,B_2})$ and $w_{G,B'_2}$ is a proper subterm of $w_{G,B_2}$. Hence, by Lemma 3.15, $\sigma(F_1) = w_{G',F_1}$. As before, we can apply Lemma 3.13 and conclude that $\sigma$ is a solution of $\langle \Delta', G', \Gamma' \rangle$.

Finally, assume that the applied rule is an *unfolding* rule. Note that the application of an *unfolding* rule does not modify the set of restrictions $\Gamma$ nor the grammar $G$. $\Delta'$ is obtained by replacing the left-hand side of an equation $u \doteq v$ in $\Delta$ by a new equation $u' \doteq v$. Since $G = G'$, it holds that $w_{G,u} = w_{G',u}$. Moreover, since $\sigma$ is a solution of $\langle \Delta, G, \Gamma \rangle$, it satisfies $\sigma(w_{G,u}) = w_{G,v}$. Hence, it suffices to check that $w_{G,u} = w_{G,u'}$ when $u \doteq v$ is replaced by $u' \doteq v$ by the rule application. But this is direct from the fact that this replacements are due to a rule application of $G$, and we are done. □

PROPOSITION 3.17. *For every initial triple $\langle \Delta_0, G_0, \Gamma_0 = \emptyset \rangle$, the determinized algorithm will compute a complete set of solved triples $\langle \Delta_1, G_1, \Gamma_1 \rangle, \ldots, \langle \Delta_n, G_n, \Gamma_n \rangle$, such that $\sigma$ is a solution of $\langle \Delta_0, G_0, \Gamma_0 = \emptyset \rangle$ iff it is a solution of some $\langle \Delta_i, G_i, \Gamma_i \rangle$, for $i \in [1, n]$.*

PROOF. Termination holds, see the argumentation on the complexity in the next section. Since we have proved soundness and completeness, it remains to show that if some intermediate $\langle \Delta, G, \Gamma \rangle$ is not solved, then an inference rule can be applied. The k-CMD algorithm represents instantiations of variables by transforming them into non-terminals of the STG. Hence, the fact that a triple $\langle \Delta, G, \Gamma \rangle$ is not solved means that there are occurrences of terminals of $G$ representing first order variables or context variables in $\Delta$.

Assume that no inference rule can be applied. We will deduce the form of the equations $u \doteq B \in \Delta$ under this assumption until we reach a contradiction. Let $A, A_1, \ldots, A_m, B, B_1, \ldots, B_m$ be term non-terminals of $G$, let $C_i, C$ be context non-terminals of $G$, and let $f, g$ be terminals of $G$ representing function symbols of arity $m$ and $m'$, respectively.

Note that $u$ cannot be of the form $f(A_1, \ldots, A_m)$ nor $f(A_1, \ldots, A_{i-1}, C_i A, A_{i+1}, \ldots, A_m)$ since, as $v$ is a non-terminal $B$ with rule $B \to g(B_1, \ldots, B_{m'})$, either DECOMPOSE (if $f = g$), or FAIL (if $f \neq g$) would be applicable. Hence, by 3.11, at this point $u$ can be of the forms $x$, $F(A)$ or $CA$, where $x$ is a terminal of the grammar representing a first-order variable and $F$ is a terminal of the grammar representing a context variable. This implies that, if $u = x$ then ELIMX is applicable, and if $u = CA$ then UNFOLD2 is applicable. Thus, $u$ can only be of the form $F(A)$. Hence, since we argued about an arbitrarily chosen equation $u \doteq v \in \Delta$, every equation $i$ in $\Delta$ is of the form $F_i(A_i) \doteq B_i$. Moreover, since neither ELIMF1, ELIMF2 nor ELIMF3 can be applied, for every terminal $F_i$ representing a context variable occurring in $\Delta$ there exists an equation $F_j(A_j) \doteq B_j$ in $\Delta$, such that $F_i$ is different from the terminal $F_j$, and $F_i$ occurs in $w_{A_j}$. Since the set $\Delta$ is finite, there exist equations $F_1(A_1) \doteq B_1, F_2(A_2) \doteq B_2, \ldots, F_n(A_n) \doteq B_n$ with $n \geq 2$ satisfying that $F_1$ occurs in $w_{A_2}$, $F_2$ occurs in $w_{A_3}$, $\ldots$, $F_{n-1}$ occurs in $w_{A_n}$, and $F_n$ occurs in $w_{A_1}$, and where the $F_i$'s are pairwise different. Let $i$ be such that $w_{B_i}$ has maximal height among the $w_{B_1}, \ldots, w_{B_n}$, say $i = 1$. Hence, we may take the equation $F_1(A_1) \doteq B_1$ and the equation $F_2(A_2) \doteq B_2$ and apply rule ELIMF4, which is a contradiction. □

The following example shows that the DECOMPOSE rule may have an exponential number of executions if multiple insertions of the same equation in one inference sequence is not prohibited. Hence, our algorithm must keep track of already treated equations in order to avoid this fact.

*Example* 3.18. Let $G$ be an STG defined by the following set of rules: $\{B_1 \to f(B_2, B_2), B_2 \to f(B_3, B_3), \ldots, B_i \to f(B_{i+1} B_{i+1}), \ldots, B_{n-1} \to f(B_n, B_n), B_n \to a, A_1 \to f(A_2, A_2'), A_2 \to f(A_3, A_3'), \ldots, A_i \to f(A_{i+1} A_{i+1}'), \ldots, A_{n-1} \to f(A_n, A_n'), A_n \to a, A_1' \to f(A_2, A_2'), A_2' \to f(A_3, A_3'), \ldots, A_i' \to f(A_{i+1} A_{i+1}'), \ldots, A_{n-1}' \to f(A_n, A_n'), A_n' \to a, A \to f(A_1, A_1'), B \to f(B_1, B_1)\}$
We now consider a decomposition sequence for the equation $A \doteq B$, it decomposes depth-first. Note that $G$ satisfies the assumption on an optimally compressed representation of $\texttt{restriction}(G, \{B\})$.

$$\{A \doteq B\} \implies \{f(A_1, A_1') \doteq B\}$$
$$\implies \{A_1 \doteq B_1, A_1' \doteq B_1\}$$
$$\implies \{f(A_2, A_2') \doteq B_1, A_1' \doteq B_1\}$$
$$\implies \{A_2 \doteq B_2, A_2' \doteq B_2, A_1' \doteq B_1\}$$
$$\implies \{f(A_3, A_3') \doteq B_2, A_2' \doteq B_2, A_1' \doteq B_1\}$$
$$\implies \{A_3 \doteq B_3, A_3' \doteq B_3, A_2' \doteq B_2, A_1' \doteq B_1\}$$
$$\vdots$$
$$\implies \{A_i \doteq B_i, A_i' \doteq B_i, A_{i-1}' \doteq B_{i-1}, \ldots, A_1' \doteq B_1\}$$
$$\vdots$$
$$\implies \{A_n \doteq B_n, A_n' \doteq B_i, A_{n-1}' \doteq B_{n-1}, \ldots, A_1' \doteq B_1\}$$
$$\implies \{a \doteq a, a \doteq a, A_{n-1}' \doteq B_{n-1}, \ldots, A_1' \doteq B_1\}$$

Hence, the depth-first strategy may lead to an exponentially long sequence of decompositions.

### 3.6 Complexity of the k-CMD Algorithm

Let $\langle \Delta = \{A_{s_1} \doteq A_{t_1}, \ldots, A_{s_n} \doteq A_{t_n}\}, G = (\mathcal{TN}, \mathcal{CN}, \Sigma, R), \Gamma = \emptyset \rangle$ be the *initial configuration* of the execution, and let $\langle \Delta', G', \Gamma' \rangle$ be the last one. Recall that $L = \max_{1 \leq i \leq n}(\texttt{height}(w_{G, A_{t_i}}))$ and $k$ denotes the number of different context variables in the problem. Let $V$ denote the set of first-order variables.

Our inference rules may add new non-terminals and their corresponding rules to the grammar. Concretely, at most $|V|$ rules of the form $x \to A$ and at most $kL$ rules of the forms $C \to f(A_1, \ldots, A_{i-1}, C_i, A_{i+1}, \ldots, A_m)$ and $C \to [\cdot]$ are added to $G$ during an execution. Therefore, at any point of the execution, any right-hand side of a rule of the current STG $G''$ of the form $f(A_1, \ldots, A_m)$ is in fact a right-hand side of a rule of the initial $G$.

We count the number of different equations $u \doteq v$ that may appear during the execution. Our equations are simple with respect to the final $G'$ by Lemma 3.11. Thus, $u$ is either of the form $A$ ($|\mathcal{TN}| + |V|$ possibilities), or $f(A_1, \ldots, A_m)$ (an original right-hand side of a rule, thus $|\mathcal{TN}|$ possibilities), or $CA$ ($kL(|\mathcal{TN}|+|V|)$ possibilities), or $f(A_1, \ldots, A_{i-1}, C_i A, A_{i+1}, \ldots, A_m)$ ($kL(|\mathcal{TN}|+|V|)$ possibilities).

On the other hand, $v$ can only be a term non-terminal $A$. an original term non-terminal, thus $|\mathcal{TN}|$ possibilities.

Therefore, the total number of different equations in a branch of non-deterministic execution is $\mathcal{O}(\texttt{depth}(G)|G|^2)$. Assuming we avoid repetition of equations, this will also be the maximum number of execution steps. Each of those steps chooses an equation and applies an inference rule to it. The corresponding operations can be performed in logarithmic time with the adequate data structures. Thus, the non-deterministic execution time is $\mathcal{O}(\texttt{depth}(G)|G|^2\texttt{log}(|G|))$.

$k$ guessings over $L$ possibilities are done during the execution. Therefore, the execution time of the deterministic version of this algorithm is $\mathcal{O}((\texttt{depth}(G))^{k+1}|G|^2\texttt{log}(|G|))$.

THEOREM 3.19. *Computing all solutions (and hence deciding solvability) of an instance of the k-context matching with dags problem can be done in polynomial time. The worst case running time is $\mathcal{O}((\texttt{depth}(G))^{k+1}|G|^2\texttt{log}(|G|))$, where k is the number of context variables and $|G|$ is the size of the input dag.*

## 4. GRAMMAR CONSTRUCTIONS

For the description and analysis of efficient algorithms for context matching, first-order unification and first-order matching of STG-compressed terms we need several extension constructions of STGs. These algorithms have as suboperations finding differences in two terms and performing instantiations of context-variables and first-order variables. The difficulties are induced by the task of performing all the required operations on the compressed representation of terms.

In [Busatto et al. 2005] it was shown how to succinctly represent the preorder traversal word of a term generated by an STG using an SCFG. We reproduce this construction in Subsection 4.1 to compute an SCFG $\texttt{Pre}_G$ with non-terminals $\mathcal{P}_s$ and $\mathcal{P}_t$ generating $\texttt{pre}(s)$ and $\texttt{pre}(t)$, respectively. We also need to compute, given $\texttt{Pre}_G$, the smallest index $k$ in which $\texttt{pre}(s)$ and $\texttt{pre}(t)$ differ. In Subsection 4.2 we show how to perform this task efficiently. Our approach is based on a recent result on compressed string processing [Lifshits 2007]. As commented above, $k$ corresponds to a unique position $p \in \texttt{Pos}(s) \cap \texttt{Pos}(t)$. In Subsection 4.3, we present the procedure, given $G$ and $k$, to extend $G$ such that a new non-terminal generates $t|_p$. Avoiding the explicit calculation of $p$ refines the approach presented in previous work in STG-compressed first-order unification [Gascón et al. 2009] in order to obtain a faster algorithm.

We also need to apply substitutions once a variable is isolated. Performing a replacement of a first-order variable $x$ by a term $u$ is easily representable with STGs by simply transforming $x$ into a non-terminal $x$ of the grammar and adding rules such that $x$ generates $u$. However, since successive replacements of variables by subterms modify the initial terms, we have to show that this does not produce an exponential increase of the size of the grammar, since its depth may be doubled after each of these operations. To this end, we develop a notion of restricted depth, and show that its value is preserved during the execution, and that the size increase at each step can be bounded by this restricted depth, which is shown in Subsection 4.4.

### 4.1 Computing the preorder traversal of a term.

In [Busatto et al. 2005] it is shown how to construct, from a given STG $G$, an SCFG $\texttt{Pre}_G$ representing the preorder traversals of the terms and contexts generated by

$$
\begin{aligned}
A \to f(A_1, \ldots, A_m) &\Rightarrow \mathcal{P}_A \to f\mathcal{P}_{A_1}\ldots\mathcal{P}_{A_m} \\
A \to C_1 A_2 &\Rightarrow \mathcal{P}_A \to \mathcal{L}_{C_1}\mathcal{P}_{A_2}\mathcal{R}_{C_1} \\
A \to A_1 &\Rightarrow \mathcal{P}_A \to \mathcal{P}_{A_1} \\
C \to C_1 C_2 &\Rightarrow \begin{cases} \mathcal{L}_C \to \mathcal{L}_{C_1}\mathcal{L}_{C_2} \\ \mathcal{R}_C \to \mathcal{R}_{C_2}\mathcal{R}_{C_1} \end{cases} \\
C \to f(A_1, \ldots, A_{i-1}, C_i, A_{i+1}, \ldots, A_m) &\Rightarrow \begin{cases} \mathcal{L}_C \to f\mathcal{P}_{A_1}\ldots\mathcal{P}_{A_{i-1}}\mathcal{L}_{C_i} \\ \mathcal{R}_C \to \mathcal{R}_{C_i}\mathcal{P}_{A_{i+1}}\ldots\mathcal{P}_{A_n} \end{cases} \\
C \to [\cdot] &\Rightarrow \begin{cases} \mathcal{L}_C \to \lambda \\ \mathcal{R}_C \to \lambda \end{cases}
\end{aligned}
$$

Fig. 4. Generating the Preorder Traversal

$G$. We reproduce that construction here, presented in Figure 4 as a set of rules indicating, for each term non-terminal $A$ and its rule $A \to \alpha$ of $G$, which rule $\mathcal{P}_A \to \alpha'$ of $\text{Pre}_G$ is required in order to make a non-terminal $\mathcal{P}_A$ of $\text{Pre}_G$ satisfy $w_{\text{Pre}_G, \mathcal{P}_A} = \text{pre}(w_{G,A})$. To this end, for each context non-terminal $C$ of $G$ we also need non-terminals of $\text{Pre}_G$ generating the preorder traversal to the left of the hole ($\mathcal{L}_C$), and the preorder traversal to the right of the hole ($\mathcal{R}_C$).

It is straightforward to verify by induction on the depth of $G$ that, for every term non-terminal $A$ of $G$, the corresponding newly generated non-terminal $\mathcal{P}_A$ of $\text{Pre}_G$ generates $\text{pre}(w_A)$.

LEMMA 4.1. *Let $G$ be an STG. A SCFG $\text{Pre}_G$ of size $\mathcal{O}(|G|)$ can be constructed in time $\mathcal{O}(|G|)$ such that, for each non-terminal $N$ of $G$, there exists a non-terminal $\mathcal{P}_N$ in $\text{Pre}_G$ satisfying $w_{\text{Pre}_G, \mathcal{P}_N} = \text{pre}(w_{G,N})$.*

### 4.2 Computing the first different position of two words.

Given two non-terminals $p_1$ and $p_2$ of an SCFG $P$, we want to find the smallest index $k$ such that $w_{p_1}[k]$ and $w_{p_2}[k]$ are different. In order to solve this problem, a linear search over the generated words $w_{p_1}$ and $w_{p_2}$ is not a good idea, since their sizes may be exponentially big with respect to the size of $P$. Hence, one may be tempted to apply a binary search since prefixes are efficiently computable with SCFGs and equality is checkable in time $\mathcal{O}(|P|^3)$, which would lead to $\mathcal{O}(|P|^4)$ time complexity. However, we will use more specific information from Lifshits' work [Lifshits 2007] to obtain $\mathcal{O}(|P|^3)$ time complexity.

LEMMA 4.2. *[Lifshits 2007] Let $G$ be an SCFG. Then a data structure can be computed in time $\mathcal{O}(|G|^3)$ which allows to answer to the following question in time $\mathcal{O}(|G|)$: given two non-terminals $N_1$ and $N_2$ of $G$ and an integer value $k$, does $w_{N_1}$ occur in $w_{N_2}$ at position $k$?*

Thus, assume that the pre-computation of Lemma 4.2 has been done (in time $\mathcal{O}(|P|^3)$), and hence we can answer whether a given $w_{p_1}$ occurs in a given $w_{p_2}$ at a certain position in time $\mathcal{O}(|P|)$.

For finding the first different position between $p_1$ and $p_2$, we can assume $|w_{p_1}| \leq |w_{p_2}|$ without loss of generality. Moreover, we also assume $w_{p_1} \neq w_{p_2}[1..|w_{p_1}|]$, i.e $w_{p_1}$ is not a prefix of $w_{p_2}$. Note that this condition is necessary for the existence of a different position between $w_{p_1}$ and $w_{p_2}$, and that this will be the case when $p_1$ and $p_2$ generate the preorder traversals of different trees. Finally, we can assume

$$\texttt{index}(p_1,p_2,k',P)= \begin{cases} 1 & \text{, if } |w_{p_1}|=1 \\ \texttt{index}(p_{11},p_2,k',P) & \text{, if } (p_1 \to p_{11}p_{12}) \in P \wedge \\ & \quad w_{p_{11}} \neq w_{p_2}[(k'+1)\ldots(k'+|w_{p_{11}}|)] \\ |w_{p_{11}}|+ & \text{, if } (p_1 \to p_{11}p_{12}) \in P \wedge \\ \texttt{index}(p_{12},p_2,k'+|w_{p_{11}}|,P) & \quad w_{p_{11}} = w_{p_2}[(k'+1)\ldots(k'+|w_{p_{11}}|)] \end{cases}$$

Fig. 5. Algorithm for the Index of the First Difference

that $P$ is in Chomsky Normal Form. Note that, if this was not the case, we can force this assumption with a linear time and space transformation.

We generalize our problem to the following question: given two non-terminals $p_1$ and $p_2$ of $P$ and an integer $k'$ satisfying $k' + |w_{p_1}| \leq |w_{p_2}|$ and $w_{p_1} \neq w_{p_2}[(k'+1)..(k'+|w_{p_1}|)]$, which is the smallest $k \geq 1$ such that $w_{p_1}[k]$ is different from $w_{p_2}[k'+k]$? (Note that we recover the original question by fixing $k' = 0$).

This generalization is solved efficiently by the recursive algorithm given in Figure 5, as can be shown inductively on the depth of $p_1$. By Lemma 4.2, each call takes time $\mathcal{O}(|P|)$, and at most $\texttt{depth}(P)$ calls are executed. Thus, the most expensive part of computing the first different position of $w_{p_1}$ and $w_{p_2}$ is the pre-computation given by Lemma 4.2, that is, $\mathcal{O}(|P|^3)$.

LEMMA 4.3. *Let $P$ be an SCFG of size $n$, and let $p_1, p_2$ be non-terminals of $P$ such that $w_{p_1} \neq w_{p_2}$. The first position $k$ where $w_{p_1}$ and $w_{p_2}$ differ is computable in time $\mathcal{O}(|P|^3)$.*

### 4.3 Isolating variables

As commented in Section 2, the index $k$ from the previous subsection defines a position $p = \texttt{iPos}(t,k)$ of a term $t$ generated by an STG $G$. We show how to compute, in linear time, an extension of the STG $G$ with a non-terminal generating $t|_p$. We use the SCFG $\texttt{Pre}_G$ presented in Definition 4.1.

*Definition* 4.4. Let $G$ be an STG. Let $N$ be a non-terminal of $G$, and let $k$ be a natural number satisfying $k \leq |\texttt{Pre}(w_{G,N})|$. We recursively define $\texttt{kExt}(G,N,k)$ as an extension of $G$ as follows:

—If $k = 1$ then $\texttt{kExt}(G,N,k) = G$. In the next cases we assume $k > 1$.
—If $(N \to f(N_1,\ldots,N_{i-1},N_i,\ldots,N_m)) \in G$ and $1 + |w_{N_1}| + \ldots + |w_{N_{i-1}}| = k' < k \leq k' + |w_{N_i}|$ then $\texttt{kExt}(G,N,k) = \texttt{kExt}(G,N_i,k-k')$.
—If $(N \to C_1A_2) \in G$ and $k \leq |w_{\texttt{Pre}_G,\mathcal{L}_{C_1}}|$ then $\texttt{kExt}(G,N,k)$ includes $\texttt{kExt}(G,C_1,k)$, which contains a non-terminal $N'$ generating the subterm of $w_{G,C_1}$ at position $\texttt{iPos}(w_{G,C_1},k)$. If $N'$ is a context non-terminal then $\texttt{kExt}(G,N,k)$ additionally contains the rule $A \to N'A_2$, where $A$ is a new term non-terminal.
—If $(N \to C_1C_2) \in G$ and $k \leq |w_{\texttt{Pre}_G,\mathcal{L}_{C_1}}|$ then $\texttt{kExt}(G,N,k)$ includes $\texttt{kExt}(G,C_1,k)$, which contains a non-terminal $N'$ generating the subterm of $w_{G,C_1}$ at position $\texttt{iPos}(w_{G,C_1},k)$. If $N'$ is a context non-terminal then $\texttt{kExt}(G,N,k)$ additionally contains the rule $C \to N'C_2$, where $C$ is a new context non-terminal.

—If $(N \to C_1 N_2) \in G$ and $k' = |w_{\text{Pre}_G, \mathcal{L}_{C_1}}| < k \le |w_{\text{Pre}_G, \mathcal{L}_{C_1}}| + |w_{N_2}|$ then $\text{kExt}(G, N, k) = \text{kExt}(G, N_2, k - k')$.

—If $(N \to C_1 N_2) \in G$ and $|w_{\text{Pre}_G, \mathcal{L}_{C_1}}| + |w_{N_2}| < k$ then $\text{kExt}(G, N, k) = \text{kExt}(G, C_1, k - |w_{N_2}| + 1)$.

—If $(A \to A_1) \in G$ then $\text{kExt}(G, A, k) = \text{kExt}(G, A_1, k)$.

—In any other case $\text{kExt}(G, N, k)$ is *undefined*.

LEMMA 4.5. *Let $G$ be an STG. Let $N$ a non-terminal of $G$, and let $k$ be a natural number such that $k \le |\text{Pre}(w_{G,N})|$. Then $G$ can be extended to an STG $G'$ in time $\mathcal{O}(|G|)$ with $\mathcal{O}(\text{depth}(G))$ new non-terminals such that one of them generates the subterm of $w_{G,N}$ at position $\text{iPos}(w_{G,N}, k)$.*

PROOF. The fact that $\text{kExt}(G, N, k)$ is an extension of $G$ satisfying the statements of the lemma follows by induction on $\text{depth}(N)$, distinguishing cases according to the definition of $\text{kExt}(G, N, k)$, and applying the definition of $\text{iPos}$ from Section 2. To compute $\text{kExt}(G, N, k)$ in linear time we first build the SCFG $\text{Pre}_G$ generating the preorder traversals of the terms generated by $G$ and pre-compute the size of the term/word generated by each non-terminal in $G$ and $\text{Pre}_G$. Both operations can be done in linear time as stated in Lemma 4.1 and Lemma 2.10. Once this pre-computations are done, $\text{kExt}(G, N, k)$ can be computed by a single run over the rules of $G$, which leads to the desired time complexity. □

### 4.4 Application of substitutions and a notion of restricted depth

Recall that, when working with STGs, we represent the application of a substitution on a first-order variable $x$ by transforming $x$ into a term non-terminal and adding the necessary rules such that $x$ generates the term to which it is assigned.

When one or more substitutions of this form are applied, in general the depth of the non-terminals of $G$ might increase. In order to see that the size increase is polynomially bounded after several substitution operations when unifying, we need a new notion of depth called Vdepth, which does not increase after an application of a substitution. It allows us to bound the final size increase of $G$. The notion of Vdepth is similar to the notion of depth, but it is 0 for the non-terminals $N$ belonging to a special set $V$ satisfying the following condition.

*Definition* 4.6. Let $G = (\mathcal{TN}, \mathcal{CN}, \Sigma, R)$ be an STG, and let $V$ be a subset of $\mathcal{TN} \cup \Sigma$. We say that $V$ is a *$\lambda$-set* for $G$ if for each term non-terminal $A$ in $V$, the rule of $G$ of the form $A \to u$ is a $\lambda$-rule, i.e. $u$ is a term non-terminal.

*Definition* 4.7. Let $G = (\mathcal{TN}, \mathcal{CN}, \Sigma, R)$ be an STG and let $V$ be a $\lambda$-set for $G$. For every non-terminal $N$ of $G$, the value $\text{Vdepth}_{G,V}(N)$, denoted also as $\text{Vdepth}_V(N)$ or $\text{Vdepth}(N)$ when $G$ and/or $V$ are clear from the context, is defined as follows (recall the convention that $\max(\emptyset) = 0$).

$\text{Vdepth}(N) := 0$ for $N \in V$
$\text{Vdepth}(N) := 1 + \max\{\text{Vdepth}(N') \mid N'$ is a non-terminal occurring in $u$, where $N \to u \in G\}$, otherwise.

The Vdepth of $G$ is the maximum of the Vdepth of its non-terminals.

The idea is that $V$ contains all first-order variables, before and after converting them into term non-terminals. The following lemma is completely straightforward from the above definitions, and states that a substitution application does not modify the Vdepth provided $X \in V$ for the substitution $X \mapsto A$.

LEMMA 4.8. *Let $G$, $V$ be as in the above definition. Let $X \in V$ be a terminal of $G$ of arity $0$, and let $A$ be a term non-terminal of $G$. Let $G'$ be the STG obtained from $G$ by transforming $X$ into a term non-terminal and adding the rule $(X \to A)$. Then, for any non-terminal $N$ of $G$ it holds that $\mathtt{Vdepth}_{G'}(N) = \mathtt{Vdepth}_G(N)$.*

We also need the fact that Vdepth does not increase due to the construction of $\mathtt{kext}(G, A, k)$ from $G$. However, we first prove a more specific statement.

LEMMA 4.9. *Let $G$ be an STG, let $C$ be a context non-terminal of $G$, let $V$ be a $\lambda$-set for $G$, let $k$ be a natural number such that $w_C|_{\mathtt{iPos}(w_C, k)}$ is a context, and let $G'$ be $\mathtt{kext}(G, C, k)$.*

*Then, for every non-terminal $N$ of $G$ it holds that $\mathtt{Vdepth}_G(N) = \mathtt{Vdepth}_{G'}(N)$, and for every new non-terminal $N'$ in $G'$ and not in $G$, it holds that $\mathtt{Vdepth}_{G'}(N') \leq \mathtt{Vdepth}_G(C)$. Moreover, the number of new added non-terminals is bounded by $\mathtt{Vdepth}_G(C)$.*

PROOF. The identity $\mathtt{Vdepth}_G(N) = \mathtt{Vdepth}_{G'}(N)$ for each non-terminal $N$ of $G$ is straightforward from the fact that $\mathtt{kext}(G, C, k)$ does not change the rules for the non-terminals occurring in $G$. To prove the fact that $\mathtt{Vdepth}_{G'}(N') \leq \mathtt{Vdepth}_G(C)$ for each new non-terminal $N'$ in $G'$ and not in $G$, plus the fact that at most $\mathtt{Vdepth}_G(C)$ new non-terminals have been added, we will use induction on $\mathtt{Vdepth}_G(C)$. The base case ($\mathtt{Vdepth}_G(C) = 1$) trivially holds since, in this case, the STG $G$ is not modified (note that necessarily $k = 1$). For the induction step we distinguish cases according to the definition of $\mathtt{kExt}(G, C, k)$:

—Assume that $(C \to f(A_1, \ldots, A_{i-1}, C', \ldots, A_m)) \in G$. Note that, since $w_C|_{\mathtt{iPos}(w_C, k)}$ is a context, it holds that $1 + |w_{A_1}| + \ldots + |w_{A_{i-1}}| = k' < k \leq k' + |w_{C'}|$. In this case, $\mathtt{kext}(G, C, k) = \mathtt{kext}(G, C', k - k')$ and, since $\mathtt{Vdepth}_G(C') < \mathtt{Vdepth}_G(C)$, the lemma directly follows by induction hypothesis.

—Assume that $(C \to C_1 C_2) \in G$ and $k \leq |w_{\mathtt{Pre}_G, \mathcal{L}_{C_1}}|$. In this case, the construction of $\mathtt{kext}(G, C, k)$ is done by computing $\mathtt{kext}(G, C_1, k)$ and adding the rule $C' \to C'_1 C_2$, where $C'_1$ is the context non-terminal generating $w_{C_1}|_{\mathtt{iPos}(w_{C_1}, k)}$ and $C'$ is an additional new non-terminal. Since $\mathtt{Vdepth}_G(C_1) < \mathtt{Vdepth}_G(C)$, by induction hypothesis, it holds that for all the new non-terminals $N'$ in $G' = \mathtt{kext}(G, C_1, k)$, $\mathtt{Vdepth}_{G'}(N') \leq \mathtt{Vdepth}_G(C_1)$ and at most $\mathtt{Vdepth}_G(C_1)$ new non-terminals have been added. It follows that at most $\mathtt{Vdepth}_G(C)$ new non-terminals have been added in the construction of $\mathtt{kext}(G, C, k)$, and $\mathtt{Vdepth}_{G'}(C'_1) \leq \mathtt{Vdepth}_G(C_1)$. Moreover, since $\mathtt{Vdepth}_G(C) = 1 + \max(\mathtt{Vdepth}_G(C_1), \mathtt{Vdepth}_G(C_2))$, $\mathtt{Vdepth}_{G'}(C') = 1 + \max(\mathtt{Vdepth}_{G'}(C'_1), \mathtt{Vdepth}_{G'}(C_2))$ and $\mathtt{Vdepth}_G(C_2) = \mathtt{Vdepth}_{G'}(C_2)$, it also holds that $\mathtt{Vdepth}_{G'}(C') \leq \mathtt{Vdepth}_G(C)$.

—Assume that $(C \to C_1 C_2) \in G$ and $k' = |w_{\mathtt{Pre}_G, \mathcal{L}_{C_1}}| < k \leq |w_{\mathtt{Pre}_G, \mathcal{L}_{C_1}}| + |w_{C_2}|$. In this case, $\mathtt{kext}(G, C, k) = \mathtt{kext}(G, C_2, k - k')$ and, since $\mathtt{Vdepth}_G(C_2) < \mathtt{Vdepth}_G(C)$, the lemma directly follows by induction hypothesis.

Finally, note that the case $(C \to C_1 C_2) \in G$ and $|w_{\text{Pre}_G, \mathcal{L}_{C_1}}| + |w_{C_2}| < k$ is not possible due to the assumption that $w_C|_{\text{iPos}(w_C, k)}$ is a context. □

LEMMA 4.10. *Let $G$ be an STG, let $N$ be a non-terminal of $G$, let $V$ be a $\lambda$-set for $G$, let $k$ be a natural number satisfying $k \leq |\text{Pre}(w_{G,N})|$, and let $G'$ be $\text{kext}(G, N, k)$.*

*Then, for every non-terminal $N'$ of $G$ it holds that $\text{Vdepth}_G(N') = \text{Vdepth}_{G'}(N')$, and for every new non-terminal $N''$ in $G'$ and not in $G$, it holds that $\text{Vdepth}_{G'}(N'') \leq \text{Vdepth}(G)$. Moreover, the number of new added non-terminals is bounded by $\text{Vdepth}(G)$.*

PROOF. The identity $\text{Vdepth}_G(N') = \text{Vdepth}_{G'}(N')$ for each non-terminal $N'$ of $G$ is straightforward from the fact that $\text{kext}(G, N, k)$ does not change the rules for the non-terminals occurring in $G$. We will prove the fact that $\text{Vdepth}_{G'}(N'') \leq \text{Vdepth}(G)$ for each new non-terminal $N''$ in $G'$ and not in $G$, plus the fact that at most $\text{Vdepth}(G)$ new non-terminals have been added by induction on $\text{depth}_G(N)$. The base case ($\text{depth}(N) = 1$) trivially holds since, in this case, the STG $G$ is not modified. For the induction step we distinguish cases according to the definition of $\text{kExt}(G, N, k)$. The only interesting cases are when $(N \to C_1 A_2) \in G$ and $k \leq |w_{\text{Pre}_G, \mathcal{L}_{C_1}}|$, and when $(N \to C_1 C_2) \in G$ and $k \leq |w_{\text{Pre}_G, \mathcal{L}_{C_1}}|$. Note that these are the only cases in which the grammar might be extended with new non-terminals after the recursive call. We will solve the first one, the other is solved analogously.

Hence, assume that $(N \to C_1 A_2) \in G$ and $k \leq |w_{\text{Pre}_G, \mathcal{L}_{C_1}}|$. In this case the non-terminal $N'$ in $\text{kext}(G, C_1, k)$ generating the subterm of $w_{G, C_1}$ at position $\text{iPos}(w_{G, C_1}, k)$ is a either a term non-terminal or a context non-terminal. We will solve the two cases separately. First assume that $N'$ is a term non-terminal. In this case $\text{kext}(G, N, k)$ is constructed as $\text{kext}(G, C_1, k)$. Since $\text{Vdepth}_G(C_1) < \text{Vdepth}_G(N)$, the lemma holds by induction hypothesis in this case. On the other hand, if $N'$ is a context non-terminal, the construction of $\text{kext}(G, N, k)$ is done by computing $\text{kext}(G, C_1, k)$ and adding the rule $A \to N' A_2$, where $A$ is an additional new term non-terminal. By Lemma 4.9, for all the new non-terminals $N''$ in $\text{kext}(G, C_1, k)$ and not in $G$, $\text{Vdepth}_{G'}(N'') \leq \text{Vdepth}_G(C_1)$. Moreover, the number of new added non-terminals is bounded by $\text{Vdepth}_G(C_1)$. Hence, $\text{Vdepth}_{G'}(N') \leq \text{Vdepth}_G(C_1)$ and, since $\text{Vdepth}_G(C_1) < \text{Vdepth}_G(N)$, at most $\text{Vdepth}_G(N) \leq \text{Vdepth}(G)$ new non-terminals have been added in the construction of $\text{kext}(G, N, k)$. Furthermore, since $\text{Vdepth}_G(N) = 1 + \max(\text{Vdepth}_G(C_1), \text{Vdepth}_G(A_2))$, $\text{Vdepth}_{G'}(A) = 1 + \max(\text{Vdepth}_{G'}(N'), \text{Vdepth}_{G'}(A_2))$ and $\text{Vdepth}_G(A_2) = \text{Vdepth}_{G'}(A_2)$, it also holds that $\text{Vdepth}_{G'}(A) \leq \text{Vdepth}_G(N) \leq \text{Vdepth}(G)$. □

## 5. NP-COMPLETENESS OF CONTEXT MATCHING WITH STGS

As a complement to Theorem 3.19, we are now ready to show that context matching with STG-compressed terms is in NP-complete. NP-hardness with STGs follows from NP-hardness of the same problem without any compression (see [Schmidt-Schauß and Schulz 1998]). Hence, we just have to prove that this problem is in NP. Our goal is to be able to guess a solution of polynomial size for a given input context matching problem, and to check it efficiently. To this end, we first introduce definitions of prefix and suffix of a context and subcontext of a term as extensions of the original STG, and argue that the size of such extensions is polynomially bounded by the size of the original STG. Part of the used ideas are borrowed from [Levy et al. 2006b; 2004], but adapted to show a concrete complexity measure.

5.1 Grammar-extensions for hole path and subcontexts

*Definition* 5.1. Let $G$ be an STG. We define the SCFG $H_G$ representing the hole paths of $w_C$ for all context non-terminals $C$ as follows. For each context non-terminal $C$ of $G$ we construct a non-terminal $H_C$ of $H_G$. For each natural number $i$ between 1 and the maximum arity of the signature $\Sigma$, we construct a non-terminal $H_i$. For each rule with a context non-terminal $C$ as left-hand side, we construct one rule of $H_G$, depending on the form of the rule of $C$ in $G$, as follows.

—if $(C \to [\cdot]) \in G$, then $H_G$ contains the rule $H_C \to \lambda$.
—if $(C \to C_1 C_2) \in G$, then $H_G$ contains the rule $H_C \to H_{C_1} H_{C_2}$.
—if $(C \to f(A_1, \ldots, A_{i-1}, C_i, A_{i+1}, \ldots, A_m)) \in G$, then $H_G$ contains the rules $H_C \to H_i H_{C_i}$.

Moreover, for each $H_i$ we construct the rule $H_i \to i$.

LEMMA 5.2. *The SCFG $H_G$ can be computed for an STG $G$ in time $\mathcal{O}(|G|)$. For every context non-terminal $C \in G$, the corresponding non-terminal $H_C \in H_G$ generates $\mathtt{hp}(w_C)$. Moreover, $|H_G| \leq |G| + M$ and $depth(H_C) \leq depth(C)$ for all $C$, where $M$ is the maximum arity of the signature.*

PROOF. It is easy to prove that $w_{H_C} = \mathtt{hp}(w_C)$ as well as $depth(H_C) \leq depth(C)$ using induction on $depth(C)$. Moreover, from every rule of $G$ we produce one rule of $H_G$, and for every $i$ between 1 and $M$ we produce one rule of $H_G$, which leads to a linear time algorithm with respect to $|G|$. □

*Definition* 5.3. Let $G$ be an STG describing first-order terms and contexts, let $C$ be a context non-terminal of $G$, and let $l$ be a natural number such that $l \leq |\mathtt{hp}(w_C)|$. We define the extension $\mathtt{Pref}(G, C, l)$ of $G$ representing a prefix of $w_C$ recursively as follows.

—If $l = 0$, then $\mathtt{Pref}(G, C, l)$ contains $G$ plus the rule $C' \to [\cdot]$, where $C'$ is a new context non-terminal. In the next cases we assume $l > 0$.
—If $l = |\mathtt{hp}(w_C)|$, then $\mathtt{Pref}(G, C, l) := G$. In the next cases we assume $l < |\mathtt{hp}(w_C)|$.
—If $(C \to C_1 C_2) \in G$ and $l \geq |\mathtt{hp}(w_{C_1})|$. Then $\mathtt{Pref}(G, C, l)$ includes $\mathtt{Pref}(G, C_2, l - |\mathtt{hp}(w_{C_1})|)$, which contains a non-terminal $C_2'$ generating the prefix of $w_{C_2}$ with $|\mathtt{hp}(w_{C_2'})| = l - |\mathtt{hp}(w_{C_1})|$, plus the rule $C' \to C_1 C_2'$, where $C'$ is a new context non-terminal.
—If $(C \to C_1 C_2) \in G$ and $l < |\mathtt{hp}(w_{C_1})|$, then, we define $\mathtt{Pref}(G, C, l)$ as $\mathtt{Pref}(G, C_1, l)$.
—If $(C \to f(A_1, \ldots, A_{i-1}, C_i, A_{i+1}, \ldots, A_m)) \in G$, then $\mathtt{Pref}(G, C, l)$ includes $\mathtt{Pref}(G, C_i, l-1)$, which contains a non-terminal $C_i'$ generating the prefix of $w_{C_i}$ with $|\mathtt{hp}(w_{C_i'})| = l - 1$, plus the rule $C' \to f(A_1, \ldots, A_{i-1}, C_i', A_{i+1}, \ldots, A_m)) \in G$, where $C'$ is a new context non-terminal.

LEMMA 5.4. *Let $G$ be an STG describing first-order terms and contexts, let $C$ be a context non-terminal of $G$, and let $l$ be a natural number such that $l \leq |\mathtt{hp}(w_C)|$. Then, $\mathtt{Pref}(G, C, l)$ is an extension of $G$ computable in time $\mathcal{O}(|G|)$. It adds at most $\mathtt{depth}(C)$ non-terminals such that one of them, called $C'$, generates the prefix of $w_C$ satisfying $|\mathtt{hp}(w_{C'})| = l$. Moreover, $\mathtt{depth}(C') \leq \mathtt{depth}(C)$ and $\mathtt{depth}(\mathtt{Pref}(G, C, l)) = \mathtt{depth}(G)$.*

PROOF. The correctness of the definition of $\mathtt{Pref}(G, C, l)$, as well as $\mathtt{depth}(C') \leq \mathtt{depth}(C)$ and $\mathtt{depth}(\mathtt{Pref}(G, C, l)) = \mathtt{depth}(G)$ can be easily shown by induction on $\mathtt{depth}(C)$. With respect to time complexity, we first precompute $|\mathtt{hp}(w_C)|$ for each context non-terminal of $G$, which can be done in linear time thanks to Lemma 2.10 and Lemma 5.2. Time complexity $\mathcal{O}(|G|)$ follows from the fact that the recursive definition decreases the depth of the involved non-terminal. □

*Definition* 5.5. Let $G$ be an STG describing first-order terms and contexts, let $C$ be a context non-terminal of $G$, and $l$ a natural number such that $l \leq |\mathtt{hp}(w_C)|$. We define the extension $\mathtt{Suff}(G, C, l)$ of $G$ representing a prefix of $w_C$ as follows:

—If $l = 0$, then $\mathtt{Suff}(G, C, l) := G$. In the next cases we assume $l > 0$.
—If $l = |\mathtt{hp}(w_C)|$ then $\mathtt{Suff}(G, C, l)$ contains $G$ plus the rule $C' \to [\cdot]$, where $C'$ is a new context non-terminal. In the next cases we assume $l < |\mathtt{hp}(w_C)|$.
—If $(C \to C_1 C_2) \in G$ and $l < |\mathtt{hp}(w_{C_1})|$. Then $\mathtt{Suff}(G, C, l)$ includes $\mathtt{Suff}(G, C_1, l)$, which contains a context non-terminal $C_1'$ generating the suffix of $w_{C_1}$ with $|\mathtt{hp}(w_{C_1'})| = |\mathtt{hp}(w_{C_1})| - l$, plus the rule $C' \to C_1' C_2$, where $C'$ is a new context non-terminal.
—If $(C \to C_1 C_2) \in G$ and $l \geq |\mathtt{hp}(w_{C_1})|$, then, with $l' := l - |\mathtt{hp}(w_{C_1})|$, we define $\mathtt{Suff}(G, C, l)$ as $\mathtt{Suff}(G, C_2, l')$.
—If $(C \to f(A_1, \ldots, A_{i-1}, C_i, A_{i+1}, \ldots, A_m)) \in G$, then we define $\mathtt{Suff}(G, C, l)$ as $\mathtt{Suff}(G, C_i, l - 1)$.

LEMMA 5.6. *Let $G$ be an STG describing first-order terms and contexts. Let $C$ be a context non-terminal of $G$, and $l$ a natural number such that $l \leq |\mathtt{hp}(w_C)|$. Then, $\mathtt{Suff}(G, C, l)$ is an extension of $G$ computable in time $\mathcal{O}(|G|)$. It adds at most $\mathtt{depth}(C)$ non-terminals such that one of them, called $C'$, generates the suffix of $w_C$ satisfying $|\mathtt{hp}(w_{C'})| = |\mathtt{hp}(w_C)| - l$. Moreover, $\mathtt{depth}(C') \leq \mathtt{depth}(C)$ and $\mathtt{depth}(\mathtt{Pref}(G, C, l)) = \mathtt{depth}(G)$.*

PROOF. The proof is analogous to the one of Lemma 5.4. □

*Definition* 5.7. Let $G$ be an STG generating terms and contexts, let $A$ be a term non-terminal of $G$, and let $p$ be a position in $w_A$. Then, we recursively define $\mathtt{pCon}(G, A, p)$ as an extension of $G$ representing the prefix context of $A$ with hole path $p$ as follows.

—if $A \to \alpha(A_1, \ldots, A_m) \in G$ and $p = i \cdot p'$ then $\mathtt{pCon}(G, A, p)$ includes $\mathtt{pCon}(G, A_i, p')$, which contains a non-terminal $C_i$ generating the context prefix of $w_{A_i}$ with $\mathtt{hp}(w_{C_i}) = p'$, plus the rule $C' \to \alpha(A_1, \ldots, C_i, \ldots, A_m)$, where $C'$ is a new context non-terminal.
—If $A \to A'$, then $\mathtt{pCon}(G, A, p) = \mathtt{pCon}(G, A', p)$.

—if $A \to C_1 A_2 \in G$ then $p = p_1 \cdot p_2$ where $p_1$ is the maximal common prefix of $p$ and $\text{hp}(C_1)$. We distinguish three cases:

—if $p_1 = \text{hp}(C_1)$ then $\text{pCon}(G, A, p)$ includes $\text{pCon}(G, A_2, p_2)$, which contains a non-terminal $C_2$ generating the context prefix of $w_{A_2}$ with $\text{hp}(w_{C_2}) = p_2$, plus the rule $C' \to C_1 C_2$, where $C'$ is a new context non-terminal.

—if $p_1 \prec \text{hp}(C_1)$ and $p_2 = \lambda$ then $\text{pCon}(G, A, p)$ is defined as $\text{Pref}(C_1, G, |p_1|)$.

—if $p_1 \prec \text{hp}(C_1)$ and $p_2 \neq \lambda$ then $p$ is of the form $p_1 \cdot i \cdot p_3$ and $\text{hp}(C_1)$ is of the form $p_1 \cdot k \cdot p_4$, for some positions $p_3$ and $p_4$, and some integers $i$ and $k$ satisfying $i \neq k$. We assume $i < k$, without loss of generality. Let $l_1$ and $l_4$ be $|p_1|$ and $|p_4|$, respectively.

Let $G_1$ be $\text{Pref}(G, C_1, l_1)$. The STG $G_1$ contains a context non-terminal $C_{11}$ generating the prefix of $w_{C_1}$ such that $|\text{hp}(C_{11})| = l_1$. Let $G_2$ be $\text{Suff}(G_1, C_1, |\text{hp}(C_1)| - l_4)$. The STG $G_2$ contains a context non-terminal $C_{12}$ generating the suffix of $w_{C_1}$ such that $|\text{hp}(C_{12})| = l_4$.

Let $G'_1$ be $\text{Suff}(G, C_1, l_1)$. The STG $G'_1$ contains a context non-terminal $C'_{12}$ generating the suffix of $w_{C_1}$ such that $|\text{hp}(C'_{12})| = |\text{hp}(C_1)| - l_1 = l_4 + 1$. Let $G'_2$ be $\text{Pref}(G'_1, C'_{12}, 1)$. The STG $G'_2$ contains a context non-terminal $C'_{11}$ generating the prefix of $w_{C'_{12}}$ such that $|\text{hp}(C'_{11})| = 1$.

At this point, note that $w_{G, C_1} = w_{G_2, C_{11}} w_{G'_2, C'_{11}} w_{G_2, C_{12}}$. Moreover, the rule of $C'_{11}$ in $G'_2$ is of the form $C'_{11} \to \alpha(A'_1, \ldots, A'_{k-1}, C''_{11}, A'_{k+1}, \ldots, A'_m)$, where all the $A'_i$ are term non-terminals of the original $G$, and generating the same terms as in $G$. Moreover, the rule of $C''_{11}$ in $G'_2$ is necessarily $C''_{11} \to [\cdot]$.

We define $\text{pCon}(G, A, p)$ as $\text{pCon}(G_2, A'_i, p_3)$, which contains a context non-terminal $C_3$ generating $w_{A'_i}[\cdot]_{p_3}$, plus the rules $C' \to C_{11} C_4$, $C_4 \to \alpha(A'_1, \ldots, A'_{i-1}, C_3, A'_{i+1}, \ldots, A'_{k-1}, A'_k, A'_{k+1}, \ldots, A_m)$, $A'_k \to C_{12} A_2$, where $C', C_4, A'_k$ are new non-terminals.

LEMMA 5.8. *Let $G$ be an STG describing terms and contexts, let $A$ be a non-terminal of $G$, and let $p$ be a position in $w_A$. Then, $\text{pCon}(G, A, p)$ contains at most $\text{depth}(A) * (2\text{depth}(A) + 3)$ new non-terminals such that one of them, called $C'$, generates the context prefix of $w_A$ with $\text{hp}(w_{C'}) = p$. Moreover, $\text{depth}(C') \leq 4\text{depth}(A)$.*

PROOF. The fact that $\text{pCon}(G, A, p)$ contains a context non-terminal $C'$ generating the context prefix of $w_A$ with $\text{hp}(w_{C'}) = p$ can be verified by induction on $\text{depth}(A)$ and distinguishing cases according to the definition of $\text{pCon}(G, A, p)$. As in previous constructions, $\text{pCon}(G, A, p)$ can be computed in a single run over the rules of the $G$. To show the upper bound to the size of the computed extension, it suffices to note that the worst case in this sense is when the rule of $A$ is of the form $A \to C_1 A_2$ and $\text{hp}(C_1)$ and $p$ are disjoint. In such a case, we add 3 new rules plus the new rules in $\text{Pref}(G, C_1, |p_1|)$ and $\text{Suff}(G, C_1, |p_1| + 1)$, where $p_1$ is the maximal common prefix between $\text{hp}(C_1)$ and $p$. The number of added non-terminals is bounded by $\text{depth}(A)$ for both the Pref and the Suff constructions by Lemma 5.4, and Lemma 5.6, respectively. The fact $\text{depth}(C') \leq 4\text{depth}(A)$ can be verified by induction on $\text{depth}(A)$ and using Lemma 5.4, and Lemma 5.6. □

### 5.2 NP-completeness of STG-context-matching

At this point we are ready to show that the STG-context-matching problem is in NP. However, we will first remark on how we represent the input and the solutions for this problem. An input consists on an STG $G$ and two non-terminals $A_s$ an $A_t$ of $G$. We want to decide whether there exists a substitution $\sigma$ for the first-order and context variables occurring in $w_{A_s}$ such that $\sigma(w_{A_s}) = w_{A_t}$. In the input of the algorithm, the first order and the context variables are 0-ary and 1-ary terminals of $G$, respectively. A solution $\sigma$ can be represented by another STG $G'$, where the first order and the context variables are term and context non-terminals of $G'$, respectively. That is, $\sigma(x) = w_{G',x}$ and $\sigma(F) = w_{G',F}$, for each first-order variable $x$ and context variable $F$. For proving NP inclusion, we just show that, if such $\sigma$ exists then there exists an extension $G'$ of $G$, which is polynomially bounded in the size of $G$, satisfying $w_{G',A_s} = w_{G',A_t} = w_{G,A_t}$. The fact that this equality can be checked in polynomial time follows again from Theorem 2.9.

LEMMA 5.9. *Let $G$ be an STG, and let $A_s$ and $A_t$ be term non-terminals of $G$. Let $\langle A_s, A_t, G \rangle$, be an STG-context-matching problem instance, and let $\sigma$ be a substitution such that $\sigma(w_{G,A_s}) = w_{G,A_t}$ (a solution). Then, there exists an extension $G'$ of $G$ such that $w_{G',A_s} = w_{G',A_t} = w_{G,A_t}$. Furthermore, $|G'|$ is polynomially bounded by $|G|$.*

PROOF. Let $\{x_1, \ldots, x_n\}$ and $\{F_1, \ldots, F_m\}$ be the set of first-order variables and context variables, respectively, occurring in $w_{G,A_s}$. For each first order variable $x_i$, $\sigma(x_i)$ is a subterm of $w_{A_t}$ at some position $p_i$. Thus, for each first-order variable $x_i$, we construct the STG $G'_{x_i} = (\mathcal{TN}'_{x_i}, \mathcal{CN}'_{x_i}, \Sigma'_{x_i}, R'_{x_i})$ as $\texttt{kExt}(A_t, G, \texttt{pIndex}(t, p_i))$, which contains a term non-terminal $A_{x_i}$ generating $w_{A_{x_i}} = \sigma(x_i)$. Then we convert $x_i$ into a non-terminal generating $\sigma(x_i)$ by defining $G_{x_i} = (\mathcal{TN}_{x_i}, \mathcal{CN}_{x_i}, \Sigma_{x_i}, R_{x_i})$ from $G'_{x_i}$ as $G_{x_i} = (\mathcal{TN}'_{x_i} \cup \{x_i\}, \mathcal{CN}'_{x_i}, \Sigma'_{x_i} - \{x_i\}, R'_{x_i} \cup \{x_i \to A_{x_i}\})$. Similarly, for each context variable $F_j$, $\sigma(F_j) = C$ is a prefix context of some subterm of $t = w_{A_t}$. Therefore, there exist positions $q_j, q'_j$ satisfying that $C$ is the prefix context of $t|_{q_j}$ with the hole at position $q'_j$. Thus, for each context variable $F_j$, we construct $G'_{F_j} = (\mathcal{TN}'_{F_j}, \mathcal{CN}'_{F_j}, \Sigma'_{F_j}, R'_{F_j})$ as $\texttt{pCon}(\texttt{kExt}(A_s, G, \texttt{pIndex}(t, q_j)), A_{F_j}, q'_j)$, where $\texttt{kExt}(A_s, G, \texttt{pIndex}(t, q_j))$ contains a term non-terminal $A_{F_j}$ generating $t|_{q_j}$, and $G'_{F_j}$ contains a context non-terminal $C_{F_j}$ generating $\sigma(F_j)$. Then we convert $F_j$ into a context non-terminal by defining $G_{F_j} = (\mathcal{TN}_{F_j}, \mathcal{CN}_{F_j}, \Sigma_{F_j}, R_{F_j})$ from $G'_{F_j}$ as $G_{F_j} = (\mathcal{TN}'_{F_j}, \mathcal{CN}'_{F_j} \cup \{F_j\}, \Sigma'_{F_j} - \{F_j\}, R'_{F_j} \cup \{F_j \to C_{F_j}\})$. Note that each extension of $G$ that instantiates certain variable is independent from the others, since all of them ask for subterms/subcontexts of $w_{A_t}$, which is ground, and does not change after substituting a variable. Hence, each $w_{A_{x_i}}$ and each $w_{C_{F_j}}$ can be defined independently from the rest using the STG $G$ given as input. In fact, without loss of generality, we can assume that the new added non-terminals for each $G_{x_i}$ and each $G_{F_j}$ are disjoint. Thus, we construct $G'$ as $G' = (\bigcup_{i=1}^{n} \mathcal{TN}_{x_i} \cup \bigcup_{j=1}^{m} \mathcal{TN}_{F_j}, \bigcup_{i=1}^{n} \mathcal{CN}_{x_i} \cup \bigcup_{j=1}^{m} \mathcal{CN}_{F_j}, \bigcap_{i=1}^{n} \Sigma_{x_i} \cap \bigcap_{j=1}^{m} \Sigma_{F_j}, \bigcup_{i=1}^{n} R_{x_i} \cup \bigcup_{j=1}^{m} R_{F_j})$.

By Lemma 4.5, each $\texttt{kExt}(A_s, G, \texttt{pIndex}(t, p_i))$ and each $\texttt{kExt}(A_s, G, \texttt{pIndex}(t, q_j))$ has at most $\texttt{depth}(G)$ new non-terminals. By the same Lemma, each $\texttt{depth}(\texttt{kExt}(A_s, G, \texttt{pIndex}(t, p_i)))$ and each $\texttt{depth}(\texttt{kExt}(A_s, G, \texttt{pIndex}(t, q_j)))$ is bounded by $\texttt{depth}(G)$. Thus, each $G_{x_i}$ has at most $\texttt{depth}(G) + 1$ new non-terminals, each $\texttt{depth}(G_{x_i})$ is bounded by $\texttt{depth}(G) + 1$. By Lemma 5.8, each $\texttt{pCon}(\texttt{kExt}(A_s, G, \texttt{pIndex}(t, q_j)), A_{F_j}, q'_j)$ has at most $\texttt{depth}(G) * (2\texttt{depth}(G) + 3)$ new non-terminals. Thus, each $G_{F_i}$ has at most $\texttt{depth}(G) + \texttt{depth}(G) * (2\texttt{depth}(G) + 3) + 1$ new non-terminals, that is $2\,\texttt{depth}(G)^2 + 4\,\texttt{depth}(G) + 1$.

In order to count $|G'|$, by the assumption that all new added non-terminals were disjoint for each STG, we can just take the sum of their size increases with respect to $G$. Therefore, $|G'|$ is bounded by $|G| + n(\texttt{depth}(G) + 1) + m(2\,\texttt{depth}(G)^2 + 4\,\texttt{depth}(G) + 1)$. □

THEOREM 5.10. *Context matching with STGs is in NP and hence it is NP-complete.*

PROOF. Let $G$ be an STG, and let $A_s$ and $A_t$ be term non-terminals of $G$. In order to verify that a given extension $G'$ of $G$ represents a solution for the match equation $\{A_s \doteq A_t, G\}$ it suffices to decide whether $w_{G',A_s} = w_{G',A_t}$ which can be done in polynomial time w.r.t $|G'|$ by Theorem 2.9. By Lemma 5.9 if $\{A_s \doteq A_t, G\}$ has a solution $\sigma$ then there exists an extension of polynomial size w.r.t $|G|$ representing $\sigma$. Thus there is a polynomial time verifier for the STG-context-matching problem, and hence it belongs to NP. Since context matching is already known to be NP-hard [Schmidt-Schauß and Schulz 1998] we obtain NP-completeness. □

For the special case of matching of strings compressed with SCFGs we obtain also NP-completeness: An instance of the matching problem for strings is a list of equations $s_1 \doteq t_1, \ldots, s_n \doteq t_n$, where $s_i, t_i$ are strings, only $s_i$ may contain string variables, and a solution $\sigma$ may replace string variables by strings, and must solve all equations, i.e. $\sigma(s_i) = t_i$ for all $i$.

COROLLARY 5.11. *String-matching where left and right hand sides are compressed using an SCFG, is NP-complete.*

PROOF. It is well-known that string matching is NP-hard [Benanav et al. 1985], and using a monadic signature, Theorem 5.10 shows the claim. □

## 6. FIRST-ORDER UNIFICATION WITH STGS

In this section we prove that the first-order unification problem can be solved in polynomial time even when the input is compressed using STGs. We will use the algorithms and constructions in Section 4, where the polynomial running time of certain constructions there is now relevant.

*Definition* 6.1. The *first-order unification problem with STG* has an STG $G$ representing first-order terms and contexts as input, plus two term non-terminals $A_s$ and $A_t$ of $G$ representing terms $s = w_{G,A_s}$ and $t = w_{G,A_t}$. Its decisional version asks whether $s$ and $t$ are unifiable. In the affirmative case, its computational version asks for a representation of the most general unifier.

Our algorithm generates the most general unifier in polynomial time and represents it again with an STG. It will make heavy use of the grammar-constructions in Section 4.

```
Input:  An STG G and term non-terminals A_s and A_t.
        (we write s and t for w_{A_s} and w_{A_t}).
While s and t are different do:
   Look for the first position k such that pre(s)[k] ≠ pre(t)[k].
   If both pre(s)[k] and pre(t)[k] are function symbols; Then
      Halt stating that the initial s and t are not unifiable
   // Here, either pre(s)[k] or pre(t)[k], say pre(s)[k], is a variable x.
   If x occurs in t|_p, where p = iPos(t,k), Then
      Halt stating that the initial s and t are not unifiable
   Extend G by the assignment {x ↦ t|_p}
EndWhile
Halt stating that the initial s and t are unifiable
```

Fig. 6. Unification Algorithm of STG-Compressed Terms

### 6.1 Outline of the algorithm

Our unification algorithm for compressed terms in Fig. 6 is a variant of Robinson's algorithm [Robinson 1965]: Given an STG $G$ as a compressed representation of two terms $s$ and $t$, we compute a smallest index $k$ in which $\texttt{pre}(s)$ and $\texttt{pre}(t)$ differ. At this point, if both $\texttt{pre}(s)[k]$ and $\texttt{pre}(t)[k]$ are function symbols, we terminate stating non-unifiability. Otherwise, either $\texttt{pre}(s)$ or $\texttt{pre}(t)$, say $\texttt{pre}(s)$, contains a variable $x$ at $k$. Note that, since the arity of the terminals in $G$ is fixed, the index $k$ corresponds to a unique position $p \in \texttt{Pos}(s) \cap \texttt{Pos}(t)$, as explained in Section 2. If $x$ properly occurs in the subterm of $t$ at $p$, then we terminate, again stating non-unifiability. Otherwise, we replace $x$ by the subterm of $t$ at $p$ everywhere, and repeat the process until both $s$ and $t$ become equal, in which case we state unifiability.

### 6.2 A polynomial time algorithm for first-order unification with STGs

From a high level perspective the structure of our algorithm described in Subsection 6.1 is very simple and rather standard: it is very much like the Robinson unification algorithm [Robinson 1965]. Many algorithms for first-order unification are variants of this scheme. They represent the terms with directed acyclic graphs (dags), implemented somehow, in order to avoid the space explosion due to the repeated instantiation of variables by terms. For example, the Martelli-Montanari-algorithm represents instantiations by equations [Martelli and Montanari 1982; Baader and Snyder 2001b]. In our setting, those terms are represented by STGs. In fact, the input is an STG $G$, and two term non-terminals $A_s$ and $A_t$ representing $s$ and $t$, respectively. In previous sections we showed how to efficiently perform all the required operations on STGs: Decide whether $s$ and $t$ are equal, generate a compressed representation for $\texttt{pre}(s)$ and $\texttt{pre}(t)$, look for the smallest index $k$ such that $\texttt{pre}(s)[k] \neq \texttt{pre}(s)[k]$, construct the term $t|_p$, where $p = \texttt{iPos}(t,k)$, and instantiate the variable $x = s|_p$ by $t|_p$.

The algorithm runs in polynomial time due to the following observations. Let $n$ and $m$ be the initial value of $\texttt{depth}(G)$ and $|G|$, respectively. We define $V$ to be the set of all the first-order variables at the start of the execution (before any of them has been converted into a non-terminal). Hence, at this point $\texttt{Vdepth}(G) = n$. The value $\mathtt{Vdepth}(G)$ is preserved to be $n$ along the execution of the algorithm thanks to Lemmas 4.8 and 4.10. Moreover, by Lemma 4.10, at most $n$ new non-terminals are added at each step. Since at most $|V|$ steps are executed, the final size of $G$ is bounded by $m + |V|n$. Each execution step takes time at most $\mathcal{O}(|G|^3)$. Thus we have proved:

THEOREM 6.2. *First-order unification of two terms represented by an STG can be done in polynomial time $\mathcal{O}(|V|(m + |V|n)^3)$, where $m$ represents the size of the input STG, $n$ represents the depth, and $V$ represents the set of different first-order variables occurring in the input terms). This holds for the decision question, as well as for the computation of the most general unifier, whose components are represented by the final STG.*

## 7. FIRST-ORDER MATCHING WITH STGS

In this section we prove that the first-order matching problem can be solved in polynomial time even when the input is compressed using STGs.

*Definition* 7.1. The *first-order matching problem with STG* has an STG $G$ representing first-order terms and contexts as input, plus two term non-terminals $A_s$ and $A_t$ of $G$ representing terms $s = w_{G,A_s}$ and $t = w_{G,A_t}$, where $t$ is ground. Its decisional version asks for the existence of a substitution $\sigma$ such that $\sigma(s) = t$ whereas its computational version asks for a representation of $\sigma$.

First-order matching is a particular case of first-order unification. However, taking advantage of the fact that one of the terms is ground leads to a faster algorithm with respect to the one presented in the previous section. We also improve previous complexity results for this problem [Gascón et al. 2008].

### 7.1 Outline of the algorithm

The structure of our algorithm is sketched in Figure 7. As commented above, the input of the problem consists of an STG $G$ as a compressed representation of two terms $s$ and $t$. As in the first-order unification case, the algorithm works with representations of the preorder traversal words of the terms $s$ and $t$ to be matched. Hence, we first compute a representation of $\mathtt{pre}(s)$ and $\mathtt{pre}(t)$. Then we find the index $k$ of the first occurrence of a variable $x$ in $\mathtt{pre}(s)$, and, given $G$ and $k$, compute $t' = t|_{\mathtt{iPos}(t,k)}$. If $t'$ is undefined we halt giving a negative answer. Otherwise we apply the substitution $\{x \to t'\}(s)$ and restart the process until all variables are replaced. Finally, let $s'$ be the term obtained from $s$ after all replacements are done. We check whether $s'$ and $t$ are syntactically equal and answer accordingly. Note that, in contrast to unification algorithm, we look for the first occurrence of a variable in $\mathtt{pre}(s)$ instead of looking for the first difference between $\mathtt{pre}(s)$ and $\mathtt{pre}(t)$. This refines the approach used in the previous section for the unification general case of first-order unification and improves time complexity results in previous work on first-order matching with STGs [Gascón et al. 2008].

In the previous section we already showed how to compute a succinct representation of $\mathtt{pre}(s)$ and $\mathtt{pre}(t)$, to compute, given a natural number $k$, the subterm of a term $t$ at position $iPos(t, k)$, and to apply a substitution. Hence, it only remains to show how to compute $k$, the index of the first occurrence of a variable in $\mathtt{pre}(s)$.

```
Input:  An STG G and term non-terminals A_s and A_t.
        (we write s and t for w_{A_s} and w_{A_t}
        and X for the set of variables in s).
Repeat |X| times:
   Look for the smallest index k such that pre(s)[k] = x ∈ X.
   If iPos(t,k) is undefined Then Halt stating that the initial s and t match.
   Extend G by the assignment {x ↦ t|_p}, where p = iPos(t,k).
EndRepeat
   If s = t Then Halt stating that the initial s and t match.
   Else Halt stating that the initial s and t do not match.
```

Fig. 7.   Matching Algorithm for STG-Compressed Terms

### 7.2  Finding the first occurrence of a variable

The task of finding the index of the first occurrence of a variable in a compressed word can be performed efficiently as stated in the following Lemma.

LEMMA 7.2. *Let $P$ be an SCFG, and let $p$ be a non-terminal of $P$ representing the preorder traversal word of a first-order term. Then, the smallest index $k$ such that $w_p[k]$ is a terminal and a variable can be computed in time $\mathcal{O}(|P|)$.*

PROOF. Let $\mathcal{X}$ denote the set of first-order variables. We define $k = \texttt{index}(p, P)$ as follows:

$$\texttt{index}(p,P) = \begin{cases} 1 & \text{, if } p \to \alpha \in P \land \alpha \in \mathcal{X} \\ \texttt{index}(p_1, P) & \text{, if } (p \to p_1 p_2) \in P \land \\ & \quad \exists x \in \mathcal{X} : x \text{ occurs in } w_{P,p_1} \\ |w_{P,p_1}| + \texttt{index}(X_2, P) & \text{, Otherwise.} \end{cases}$$

Note that we assumed that $P$ is in Chomsky Normal Form. If this was not the case, we can force this assumption with a linear time and space transformation. The fact that $\texttt{index}(p, P)$ computes the smallest index $k$ such that $w_p[k]$ is a variable can be shown by induction on $\texttt{depth}(p)$. With respect to the time complexity, for each non-terminal $p$ of an SCFG $P$, both the number $|w_p|$ and whether $w_p$ contains a variable can be precomputed in linear time as stated in Lemmas 2.10 and 2.11, respectively. When these pre-computations are done, $\texttt{index}(p, P)$ can be computed by a single run over the rules of $P$ and hence, it runs also in linear time.  □

### 7.3  A polynomial time algorithm for first-order matching with STGs

The algorithm presented in the previous section runs in polynomial time due to the following observations. Let $n$ and $m$ be the initial value of $\texttt{depth}(G)$ and $|G|$, respectively. We define $V := \mathcal{X}$ to be the set of all the first-order variables at the start of the execution (before any of them has been converted into a non-terminal). As in the unification case, the final size of the grammar is bounded by $m + |V|n$ thanks to Lemmas 4.8 and 4.10. Our algorithms iterates at most $V$ times. By Lemmas 4.5, and 7.2 each iteration takes linear time. Finally we check equality of two words generated by an SCFG $P$, which takes time $\mathcal{O}(|P|^3)$ thanks to Theorem 2.9. Hence, we have the following:

Theorem 7.3. *First-order matching of two terms represented by an STG can be done in polynomial time $\mathcal{O}((m + |V|n)^3)$, where m represents the size of the inputted STG, n represents its depth, and V represents the set of different first-order variables occurring in the inputted terms). This holds for the decision question, as well as for the computation of the unifier, whose components are represented by the final STG.*

## 8. CONCLUSION AND FURTHER WORK

We analyzed the complexity of context matching under different representations of terms like dags and STGs. Regarding the term compression using STGs, we showed that the context matching problem with STGs is NP-complete. Furthermore, we presented instantiation-based algorithms for the first-order matching problem and the first-order unification problem, that can be immediately executed on the compressed representation of large terms and run in polynomial time on the size of the representation. It would be interesting to investigate optimizations for these algorithms, as well as finding an improved upper bound. We also believe that it would be natural to consider the context matching problem using an STG encoding for terms under certain restrictions like fixing the number of context variables. In this sense we believe that our techniques could be useful to show that the one context unification problem is in NP when the input is represented by an STG. This problem has been solved for plain terms as input in [Gascón et al. 2008].

For the dag representation we found a polynomial context matching algorithm for the case where the number of context variables is fixed. Since the problem of context matching is NP-complete, this result is interesting because it closely links a complexity jump to a specific restriction on the original problem.

From a more general point of view, we believe that it would be interesting to investigate other variations of context unification and matching problems. Modifications such as allowing several holes in a context add expressiveness to the problem in order to encode complex questions about terms. It is also important to reconsider the complexity of sets of solutions for this variations when using different representations of terms.